\documentclass[aps,prb,twocolumn,showpacs,superscriptaddress,floatfix]{revtex4}
\usepackage{bm}
\usepackage{color}
\usepackage{amssymb}
\usepackage{graphicx}
\usepackage{amsmath}
\usepackage{dcolumn}
\usepackage{bm}
\usepackage[german,english]{babel}

\newcommand{\bq}{\mathbf{q}}

\newcommand{\beq}{\begin{eqnarray}}
\newcommand{\eeq}{\end{eqnarray}}
\newcommand{\be}{\begin{equation}}
\newcommand{\ee}{\end{equation}}

\begin{document}

\title{Screening and Collective Modes in Disordered Graphene Antidot Lattices%
}
\author{Shengjun Yuan}
\email{s.yuan@science.ru.nl}
\affiliation{Radboud University of Nijmegen, Institute for Molecules and Materials,
Heijendaalseweg 135, 6525 AJ Nijmegen, The Netherlands}
\author{Fengping Jin}
\affiliation{Institute for Advanced Simulation, Julich Supercomputing Centre, Research
Centre Julich, D-52425 Julich, Germany}
\author{Rafael Rold\'an}
\email{rroldan@icmm.csic.es}
\affiliation{Instituto de Ciencia de Materiales de Madrid, CSIC, Cantoblanco E28049
Madrid, Spain}
\author{Antti-Pekka Jauho}
\affiliation{Center for Nanostructured Graphene (CNG), DTU Nanotech, Department of Micro-
and Nanotechnology, Technical University of Denmark, DK-2800 Kongens Lyngby,
Denmark}
\author{M. I. Katsnelson}
\affiliation{Radboud University of Nijmegen, Institute for Molecules and Materials,
Heijendaalseweg 135, 6525 AJ Nijmegen, The Netherlands}
\date{\today }

\begin{abstract}
The excitation spectrum and the collective modes of graphene antidot
lattices (GALs) are studied in the context of a $\pi$-band
tight-binding model. The dynamical polarizability and dielectric function
are calculated within the random phase approximation. The effect of different
kinds of disorder, such as geometric and chemical disorder, are included in
our calculations. We highlight the main differences of GALs with respect to
single-layer graphene (SLG). Our results show that, in addition to the
well-understood bulk plasmon in doped samples, inter-band plasmons appear in
GALs. We further show that the static screening properties of undoped and
doped GALs are quantitatively different from SLG.
\end{abstract}

\maketitle

\section{Introduction}

Graphene antidot lattices (GAL) -- regular nanoscale perforations of the
pristine graphene sheet -- offer a way of creating a band gap in graphene.
Earliest investigations~\cite{Shima1993} of these structures date in fact
back to the dark ages before the graphene era,\cite{Novoselov2004} but new
momentum was gained when Petersen et al.\cite{PP08} speculated in 2008 that
these structures may have many interesting applications, even as a platform
for spin-based quantum computation. Subsequently, scores of theoretical
papers have addressed various properties of GALs using a variety of
theoretical tools (e.g., Dirac cone approximation for
the underlying graphene spectrum,\cite{PP08} density functional theory,\cite%
{Furst2009} or within the tight-binding model\cite{YRJK13}). Rather
than attempting to review this vast literature, we merely state that in our
opinion the electronic structure and its dependence on the underlying lattice
symmetry and shape of the antidots,\cite{Petersen2011,Liu2013} as well as
transport and optical properties of perfect GALs, are fairly well understood,
\cite{Furst2009} and what remains to be investigated concerns the role of
interactions, disorder, and extension of the present theoretical methods to
systems with large unit cells, such as the ones encountered in the
laboratory. What really has made GALs interesting is the rapid development
in fabrication techniques, and today several methods exist to create
(reasonably) regular structures with periods in low tens of nanometers - a
length scale at which the created gaps are predicted to be in hundreds of
millivolts, i.e., approaching the technologically relevant numbers. A short
and incomplete catalogue of fabrication methods includes block-copolymer
masks,\cite{Kim2010,Bai2010,Kim2012} nanoimprint lithography,\cite{Liang2010} e-beam lithography (either using conventional masks\cite%
{Shen2008,Begliarbekov2011,Giesbers2012,Zhang2013} or focused e-beam direct
writing of holes\cite{Rodriguez-Manzo2009}), ion beam etching,\cite{Eroms2009} nanosphere masks,\cite{Sinitskii2010,Wang2013} or nanoporous alumina
membranes as etch masks.\cite{Tada2011,Shimizu2012,Zeng2012}

The extraordinary electronic and optical properties of graphene have recently brought a lot of attention to this material as an ideal candidate for plasmonics applications.\cite{GPN12} Plasmons, which are collective density oscillations of an electron liquid,\cite{AFS82,GV05} have been extensively studied in pristine graphene within the random phase approximation (RPA), see, e.g. Refs.~
\onlinecite{S86,WSSG06,HS07,RGF10}.  In fact, understanding the screening properties of this material is essential in order to exploit its unique properties for plasmonic devices.\cite{GPN12}  However, we are aware of only
one paper (of which one of us is a co-author of) of screening in GALs;\cite%
{Schultz2011} in that paper the polarization function and plasmons of
perfect GALs were studied in the $qa\to 0$ limit ($q$ is the wave-vector and $%
a$ is the graphene lattice constant). In the present paper we present
extensive, and complementary, results for the finite $qa$ case: using a numerical method we study
the screening properties of \textit{disordered} GALs. The density of states
(DOS) is obtained from a numerical solution of the time-dependent
Schr\"odinger equation,\cite{YRK10} and the polarization function is
calculated from the Kubo formula.\cite{YRK10,YRRK11} The dielectric function
is obtained within the RPA, using the numerically computed polarization functions. 
The presence of disorder is unavoidable in  experimental realizations of GALs. Here, we consider  the generic
kinds of disorder found in these systems, such as a random
deviation of the periodicity and fluctuations of the radii of the nanoholes
from the perfect array, as well as the effect of resonant scatterers in the
sample (such as vacancies, adatoms, etc.). We find that gapped and almost
dispersionless plasmons may exist in GALs, due to inter-band transitions
between the narrow bands characteristic of the GAL spectrum. However, these
modes are expected to be highly damped. For doped samples, the classical
plasmon mode with a dispersion relation proportional to $\sqrt{q}$ is also
present. However, we find that the dispersion of this mode differs
substantially from the corresponding dispersion of the plasmon mode in SLG.
Finally, we study the main characteristics of the static screening in GALs,
paying special attention to their differences with respect to SLG.

The paper is organized as follows. In Sec. \ref{Sec:Method} we present the
details of the method. The main characteristics of the excitation spectrum
are discussed in Sec. \ref{Sec:Dielectric}. The static dielectric screening
is discussed in Sec. \ref{Sec:Static}, and our main conclusions are
summarized in Sec. \ref{Sec:Conclusions}.

\section{Model and method}

\label{Sec:Method}

We model a disordered GAL by the real space tight-binding Hamiltonian\cite%
{YRJK13}%
\begin{equation}
\mathcal{H}=-\sum_{\langle i,j\rangle}(t_{ij}c_{i}^{\dagger }c_{j}+\mathrm{%
h.c})+\sum_{i}v_{i}c_{i}^{\dagger }c_{i}+\mathcal{H}_{imp},  \label{Eq:H0}
\end{equation}%
where $c_{i}^{\dagger }$ ($c_{i}$) creates (annihilates) an electron on site
$i$ of the honeycomb lattice of graphene, $t_{ij}$ is the nearest neighbor
hopping integral and $v_{i}$ is the on-site potential. The effect of
isolated vacancies can be modeled by setting the hopping amplitudes to other
sites to zero or, alternatively, with an on-site energy $v_{i}\rightarrow
\infty $. Further, additional resonant impurities, such as hydrogen adatoms,
can be accounted by the term $\mathcal{H}_{imp}$ in Eq. (\ref{Eq:H0}):%
\begin{equation}
\mathcal{H}_{imp}=\varepsilon
_{d}\sum_{i}d_{i}^{\dagger}d_{i}+V\sum_{i}\left( d_{i}^{\dagger}c_{i}+%
\mathrm{h.c}\right) ,  \label{Eq:Himp}
\end{equation}%
where $\varepsilon _{d}$ is the on-site potential on the ``hydrogen''
impurity and $V$ is the hopping amplitude between carbon and hydrogen atoms.%
\cite{Robinson08,WK10,YRK10} In our calculations, we fix the temperature to $%
T=300$~K and use periodic boundary conditions for both the polarization
function and the density of states. The size of the system used in our
simulations is $6600\times 6600$ atoms. Throughout this paper we ignore the
effects due to spin (in our previous work \cite{YRJK13} we give a short
discussion of this point), and thus the spin degree of freedom merely
contributes a degeneracy factor 2 and is omitted for simplicity in Eq.~(\ref%
{Eq:H0}).

\begin{figure}[t]
\begin{center}
\mbox{
\includegraphics[width=0.5\columnwidth]{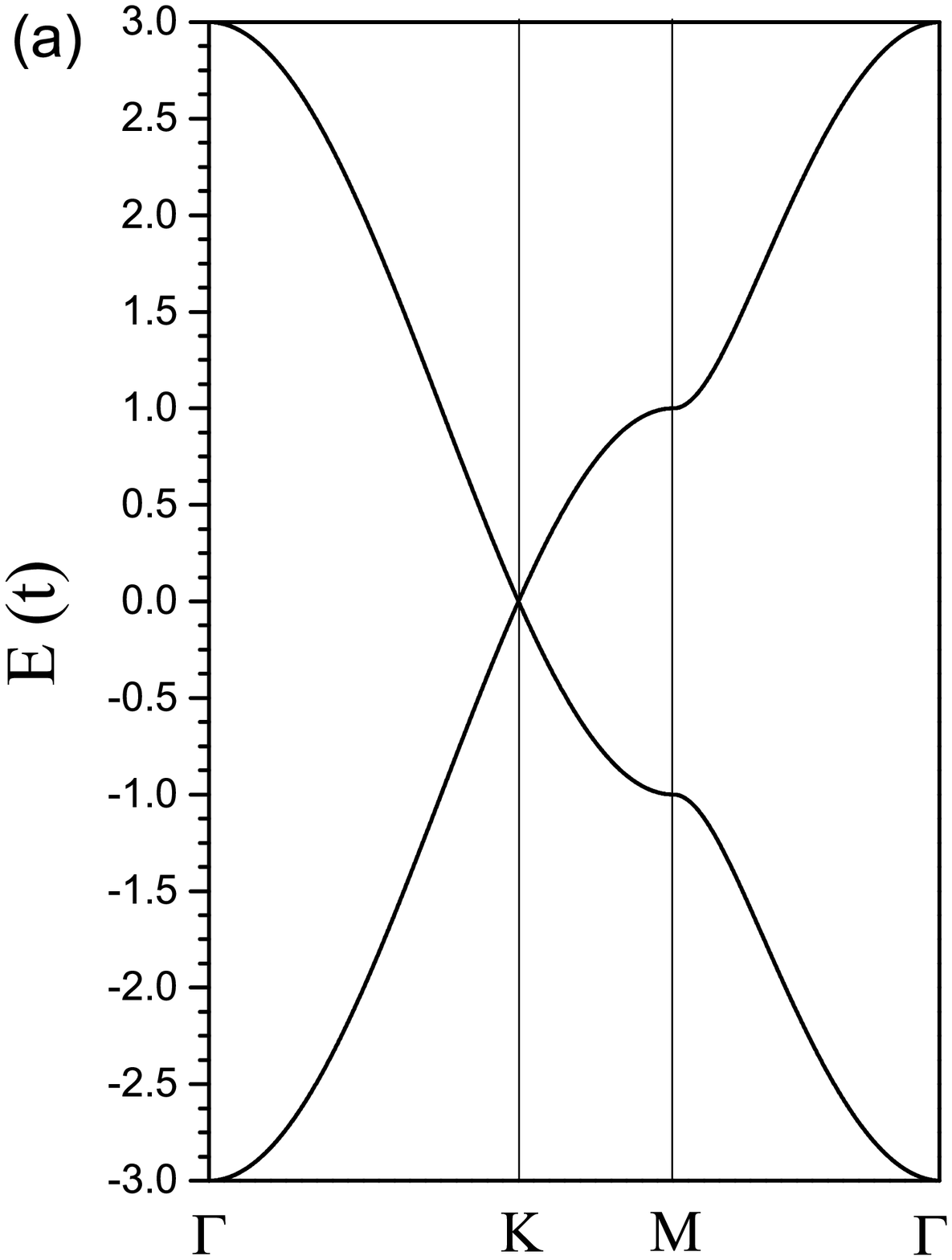}
\includegraphics[width=0.5\columnwidth]{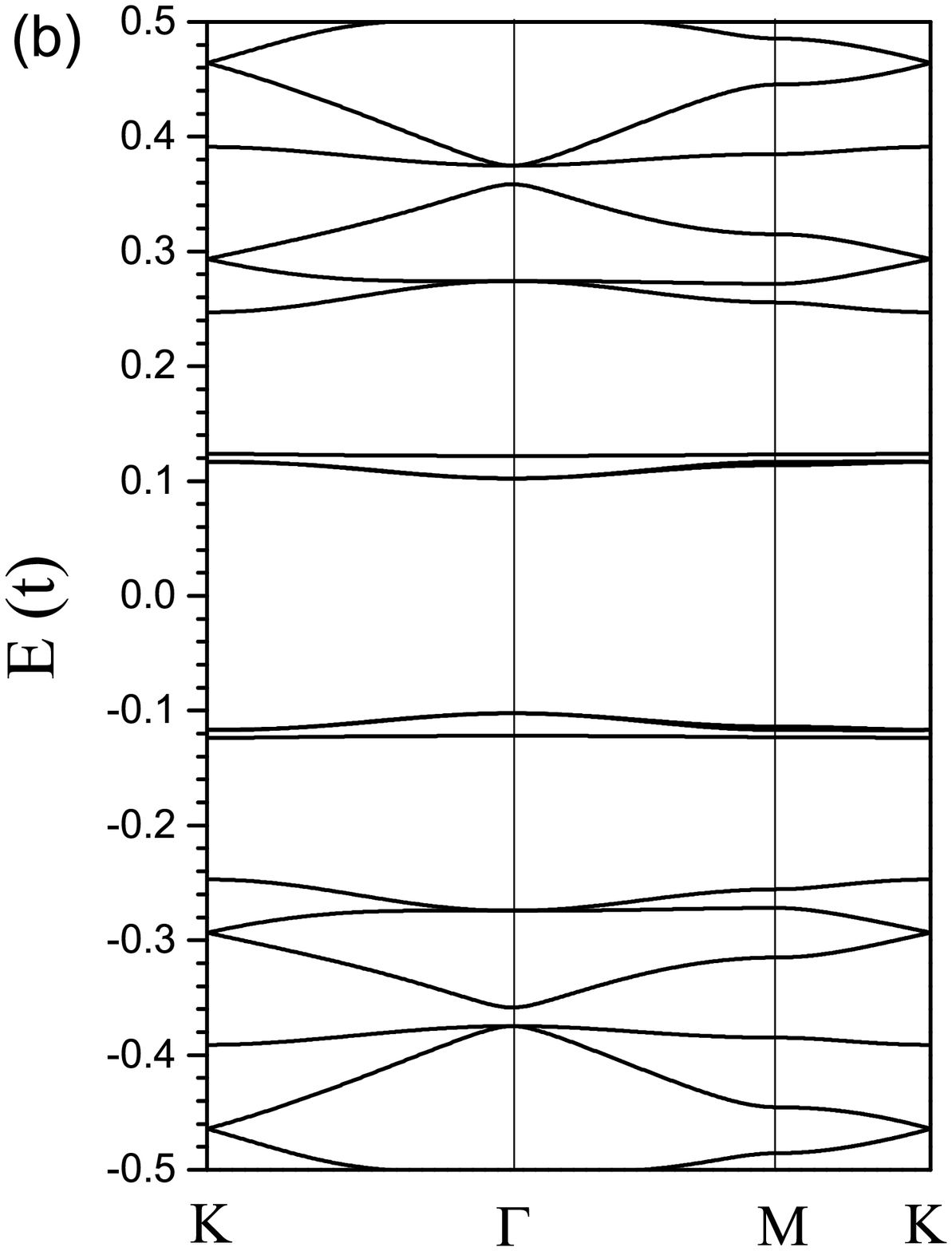}
}
\end{center}
\caption{Band structure of SLG (a) and of a $\{10,6\}$ GAL (b). }
\label{Fig:Bands}
\end{figure}

Following Refs. \onlinecite{PP08,YRJK13}, we model a GAL by creating a
hexagonal lattice of (approximately) circular holes of a given radius $R$,
and a separation $P=\sqrt{3}L$ between the centers of two consecutive holes,
where $L$ is the side length of the hexagonal unit cell (see Fig.\ref%
{Fig:notation}). GALs in this symmetry class can be labeled by the
parameters $\{L,R\}$, in units of the graphene lattice constant $a=\sqrt{3}%
\tilde{a}\approx 2.46$~\AA , where $\tilde{a}\approx 1.42$~\AA ~ is the
interatomic separation. It should be noted that there are many other
possible realizations of a GAL: both the antidot shape and its edge structure, as well as  the underlying
lattice symmetry can be varied in a number of ways. The present model is
chosen for several reasons: it is a generic model used in many previous
studies, and it displays a gapped band structure,\cite{Petersen2011,Liu2013,Foot_Gap} which is of special
interest for the present study (see Fig. \ref{Fig:Bands}). Further, to the best of our knowledge, no experiments
on plasmons in GALs have yet been reported, and in a proof-of-principle
study (such as the present one) a thorough exploration of the vast parameter
space does not seem warranted. On the other hand, when experiments emerge we
can easily extend our calculations to any particular geometry. We consider
two kinds of geometric disorder in our systems, which lead to deviations of
the GALs from perfect periodicity. First, we allow the center of the holes
to float with respect to their position in the perfect periodic lattice $(x,
y)$ around $(x \pm l_C,y \pm l_C)$. Second, we allow the radius of the holes
to randomly shrink or expand within the range $[R-r_R, R+r_R]$. All along
this paper, we express $l_C$ and $r_R$ in units of $a$.
\begin{figure}[t]
\includegraphics[width=0.8\columnwidth]{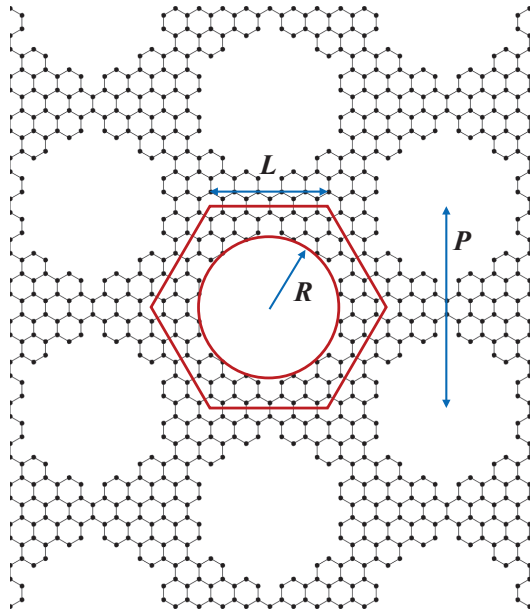}
\caption{Sketch of a GAL, with the set of geometrical parameters that are used to define it as explained in the text.}
\label{Fig:notation}
\end{figure}

The algorithm used in our numerical calculations is based on an efficient
evaluation of the time-evolution operator $e^{-i\mathcal{H}t}$ (we use units
such that $\hbar =1$) and Fermi-Dirac distribution operator $n_{F}\left(
H\right) =1/\left[ e^{\beta \left( H-\mu \right) }+1\right] $ ($\beta
=1/k_{B}T$ where $T$ is the temperature and $k_{B}$ is the Boltzmann
constant, and $\mu $ is the chemical potential) in terms of the Chebyshev
polynomial representation.\cite{RM97,YRK10} The initial state $\left\vert
\varphi \right\rangle $ is a random superposition of all the basis states in
the real space, i.e.,\cite{HR00,YRK10}
\begin{equation}
\left\vert \varphi \right\rangle =\sum_{i}a_{i}c_{i}^{\dagger }\left\vert
0\right\rangle ,  \label{Eq:phi0}
\end{equation}%
where $\left\vert 0\right\rangle $ is the electron vacuum state and $a_{i}$
are random complex numbers normalized as $\sum_{i}\left\vert
a_{i}\right\vert ^{2}=1$. Since one initial state in our calculation
contains all the eigenstates of the whole spectrum, averaging over different
initial states is not required in our numerical computations.\cite%
{HR00,YRK10} Moreover, since the system used contains millions of carbon
atoms, and one specific disorder configuration contains a large number of
different local configurations, there is no need to average over different
realizations of the disorder.\cite{YRK10} As shown in Refs. %
\onlinecite{HR00,YRK10}, the density of states (DOS) of the system can be
calculated from the Fourier transform of the overlap between the
time-evolved state $\left\vert \varphi (t)\right\rangle =e^{-i\mathcal{H}%
t}\left\vert \varphi \right\rangle $ and the initial state $\left\vert
\varphi \right\rangle $:
\begin{equation}
d \left( \varepsilon \right) =\frac{1}{2\pi }\int_{-\infty }^{\infty
}e^{i\varepsilon t}\left\langle \varphi |\varphi (t)\right\rangle dt.
\label{Eq:DOS}
\end{equation}%

Furthermore, the dynamical polarization can be obtained from the Kubo formula \cite{K57}
as%
\begin{equation}
\Pi \left( \mathbf{q},\omega \right) =\frac{i}{A}\int_{0}^{\infty }d\tau
e^{i\omega \tau}\left\langle \left[ \rho \left( \mathbf{q},\tau\right) ,\rho
\left( -\mathbf{q},0\right) \right] \right\rangle ,  \label{Eq:Kubo}
\end{equation}%
where $A$ denotes the area of the unit cell, 
$
\rho \left( \mathbf{q}\right)
=\sum_{i}c_{i}^{\dagger }c_{i}\exp \left( i\mathbf{q\cdot r}_{i}\right) 
$ 
is the density operator, and the average is taken over the canonical ensemble. For the case of the
single-particle Hamiltonian, the polarization function (\ref{Eq:Kubo}) can be written as\cite%
{YRK10}%
\begin{eqnarray}
&&\Pi \left( \mathbf{q},\omega \right) =-\frac{2}{A}\int_{0}^{\infty }d\tau
e^{i\omega \tau}  \notag  \label{Eq:Kubo2} \\
&&\times \text{Im}\left\langle \varphi \right\vert n_{F}\left( H\right)
e^{iH\tau}\rho \left( \mathbf{q}\right) e^{-iH\tau}\left[ 1-n_{F}\left(
H\right) \right] \rho \left( -\mathbf{q}\right) \left\vert \varphi
\right\rangle .  \notag \\
&&
\end{eqnarray}%
By introducing the time
evolution of two wave functions
\begin{eqnarray}
\left\vert \varphi _{1}\left(\tau\right) \right\rangle
&=&e^{-iH\tau}\left[ 1-n_{F}\left( H\right) \right] \rho \left( -\mathbf{q}%
\right) \left\vert \varphi \right\rangle , \\
\left\vert \varphi _{2}\left( \tau\right) \right\rangle
&=&e^{-iH\tau}n_{F}\left( H\right) \left\vert \varphi \right\rangle ,
\end{eqnarray}%
we obtain the real and imaginary part of the dynamical polarization as\cite{YRK11}
\begin{eqnarray}  \label{Eq:RePi-ImPi}
\text{Re}\Pi \left( \mathbf{q},\omega \right) &=&-\frac{2}{A}%
\int_{0}^{\infty }d\tau\cos (\omega \tau)~\text{Im}\left\langle \varphi
_{2}\left( \tau\right) \left\vert \rho \left( \mathbf{q}\right) \right\vert
\varphi _{1}\left( \tau\right) \right\rangle ,  \notag \\
\text{Im}\Pi \left( \mathbf{q},\omega \right) &=&-\frac{2}{A}%
\int_{0}^{\infty }d\tau\sin (\omega \tau)~\text{Im}\left\langle \varphi
_{2}\left( \tau\right) \left\vert \rho \left( \mathbf{q}\right) \right\vert
\varphi _{1}\left( \tau\right) \right\rangle .  \notag \\
&&
\end{eqnarray}%
In the numerical calculation, the integral of Eq. (\ref{Eq:Kubo}) and (\ref{Eq:RePi-ImPi}) is replaced by the sum of integrated function at finite time steps. The value of the time step is set to be $\pi/E_{\rm max}$ which is small enough as to cover the whole spectrum. Here $E_{\rm max}$ is the maximum absolute value of the energy eigenvalues, and for pristine graphene and GALs it is $3t$.  The number of time steps determines the energy resolution and typically we use 1024 time steps.  We also use a Gaussian window to alleviate the effect of the finite time used in the numerical integration in Eq. (\ref{Eq:Kubo}) and (\ref{Eq:RePi-ImPi}).

We notice here that the presence of the antidot array breaks the translational invariance of the graphene layer. Hence, in the most general case, the polarization function $\Pi$ should be a matrix in the $\bf G$, $\bf G'$ space, where $\bf G$ and $\bf G'$ are reciprocal lattice vectors associated to the new periodicity imposed in the system by the GAL.\cite{Schultz2011} Therefore, in writing the polarization function in the form of Eq. (\ref{Eq:Kubo}), we are neglecting local field effects, which is equivalent to assuming that the relevant wave-vectors $q$ are small compared to the reciprocal lattice vectors $G\propto 1/L$. This approximation is justified for the small value of $L$ considered here, $L=10a$, although local field effects should be taken into account for large values of $L$. This effect is beyond the scope of this paper, and it will be discussed in a future work.\footnote{S. Yuan {\it et al.}, in preparation.}
Furthermore, the dynamical polarization function is
anisotropic and depends on the direction of $\bq$ in the Brilloiun zone. In this paper, for simplicity, we fix the wave-vector $\mathbf{q}$ along K-$\Gamma $ direction.

The screening properties of the GAL are
determined by the dielectric function $\mathbf{\varepsilon }\left( \mathbf{q}%
,\omega \right) $, which we consider here within the RPA:
\begin{equation}
\mathbf{\varepsilon }\left( \mathbf{q},\omega \right) =\mathbf{1}-V\left(
q\right) \Pi \left( \mathbf{q},\omega \right) ,  \label{Eq:Epsilon}
\end{equation}%
where
\begin{equation}
V\left( q\right) =\frac{2\pi e^{2}}{\kappa q}
\end{equation}%
is the 2D Fourier transformation of the Coulomb interaction, and $\kappa $
is the dielectric constant of the embedding medium (we take in all our plots $\kappa=1$). Although self-energy and vertex corrections are not included within the RPA, this approximation is sufficient to capture the main features of the plasmon modes, which is the main focus of this work. From the dielectric
function, we can study the collective excitations of the system. The
dispersion relation for the collective modes is obtained from the solution
of
\begin{equation}
\mathrm{Re}~\varepsilon (\mathbf{q},\omega _{pl}(q))=0,  \label{Eq:Plasmons}
\end{equation}%
where $\omega _{pl}$ is the energy of the collective (plasmon) mode. The
condition for those modes to be long-lived is that $\mathrm{Im}~\Pi (\mathbf{%
q},\omega _{pl})=0$, such that the plasmon cannot decay into electron-hole
pairs. Otherwise, there will be a finite damping of the mode $\gamma\propto {\rm Im}~\Pi(\bq,\omega_{pl})$.

\section{Dielectric function and collective modes}

\label{Sec:Dielectric}

\begin{figure}[t]
\begin{center}
\mbox{
\includegraphics[width=0.5\columnwidth]{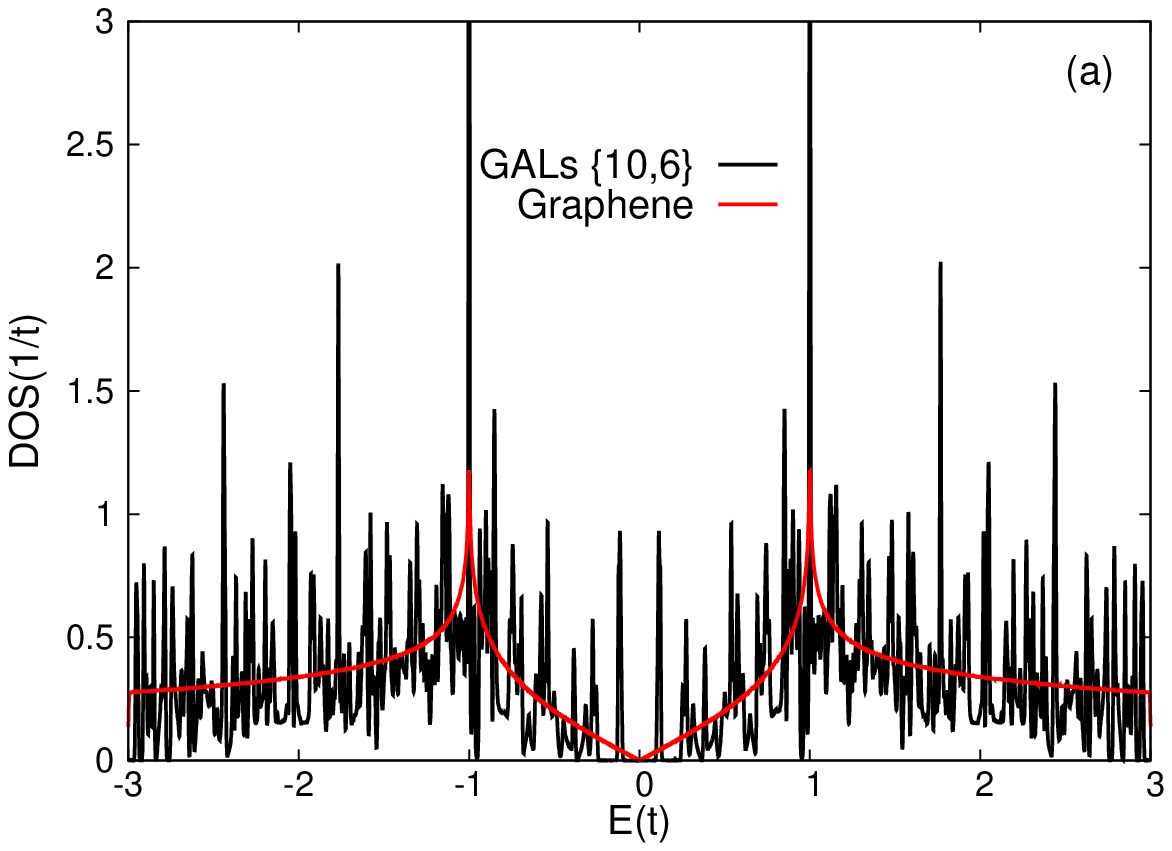}
\includegraphics[width=0.5\columnwidth]{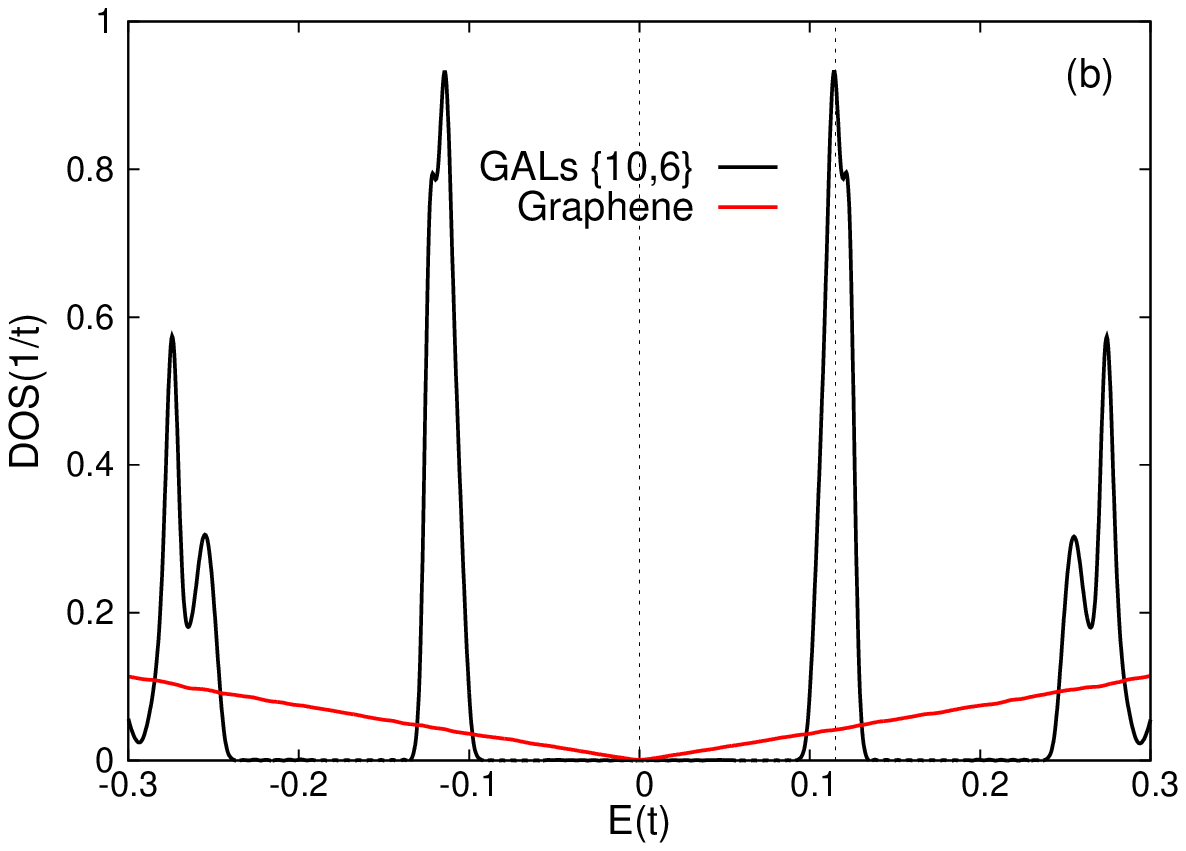}
}
\mbox{
\includegraphics[width=0.51\columnwidth]{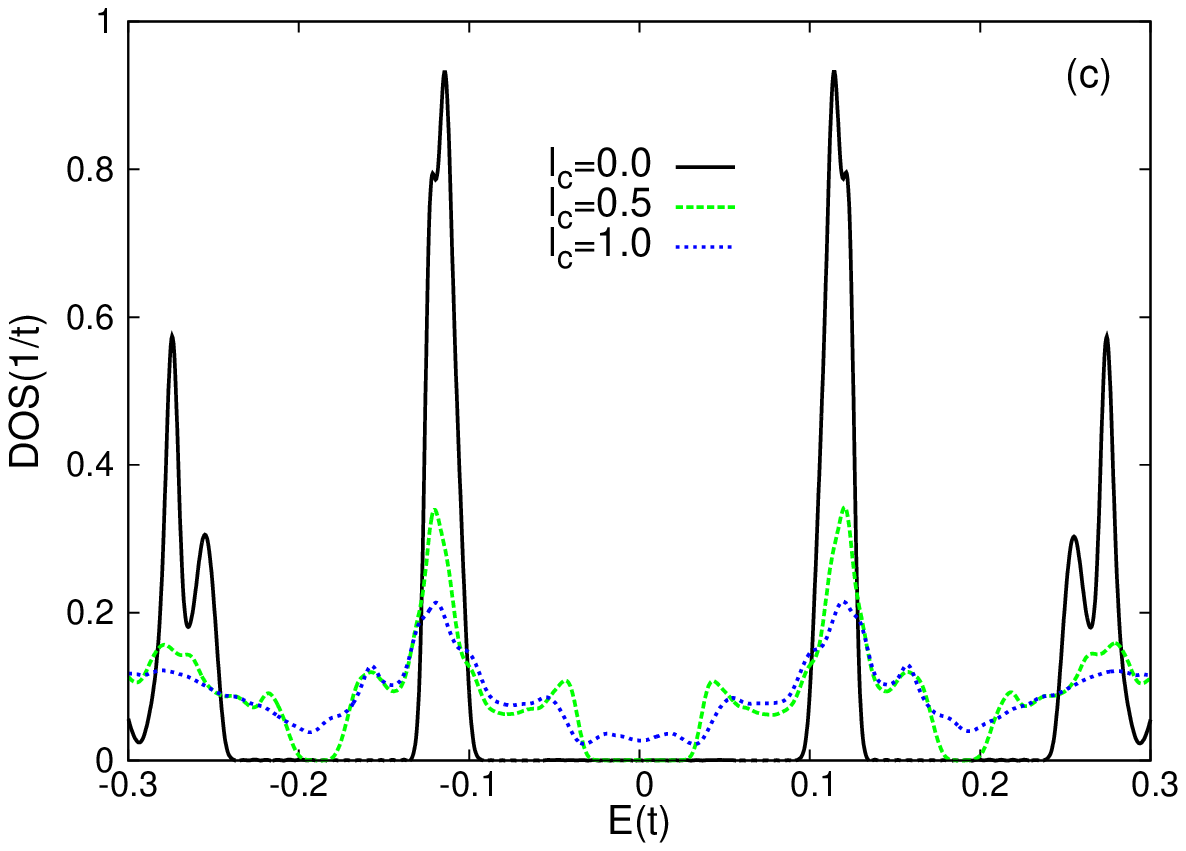}
\includegraphics[width=0.51\columnwidth]{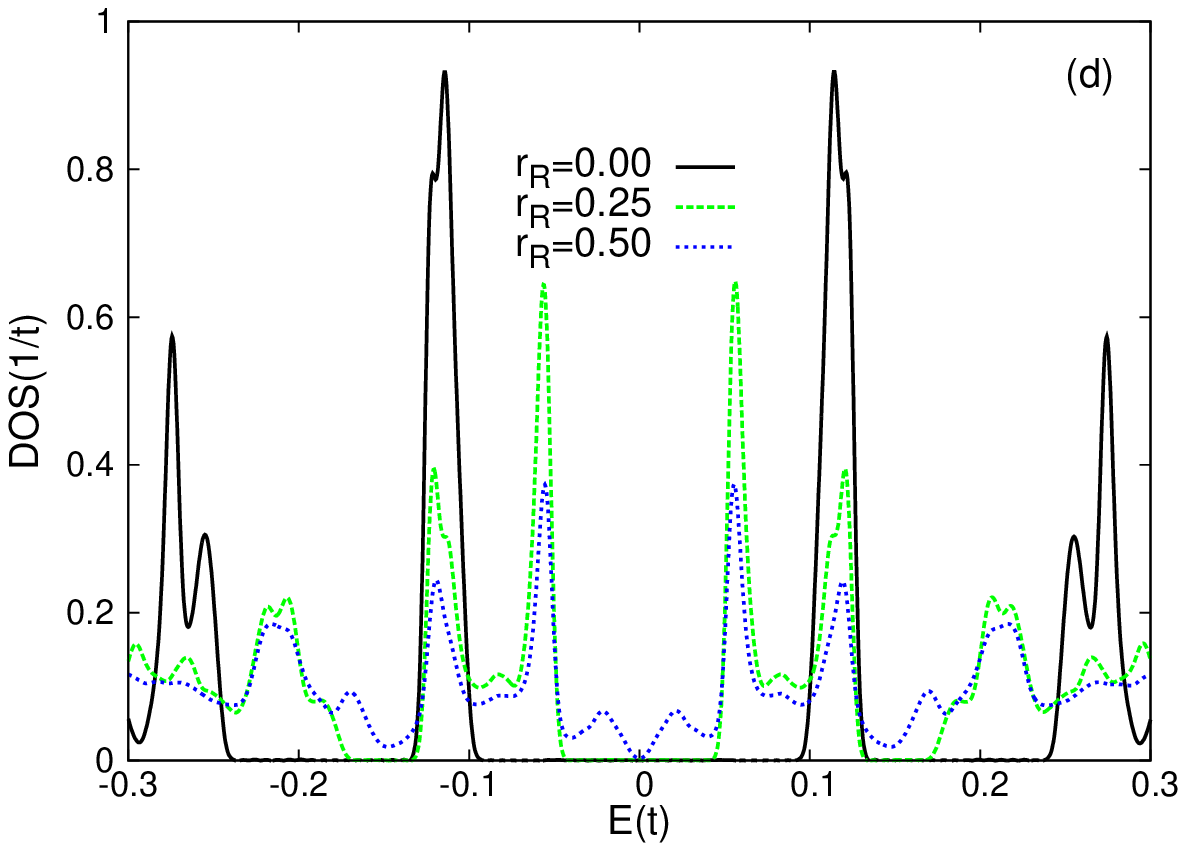}
}
\mbox{
\includegraphics[width=0.51\columnwidth]{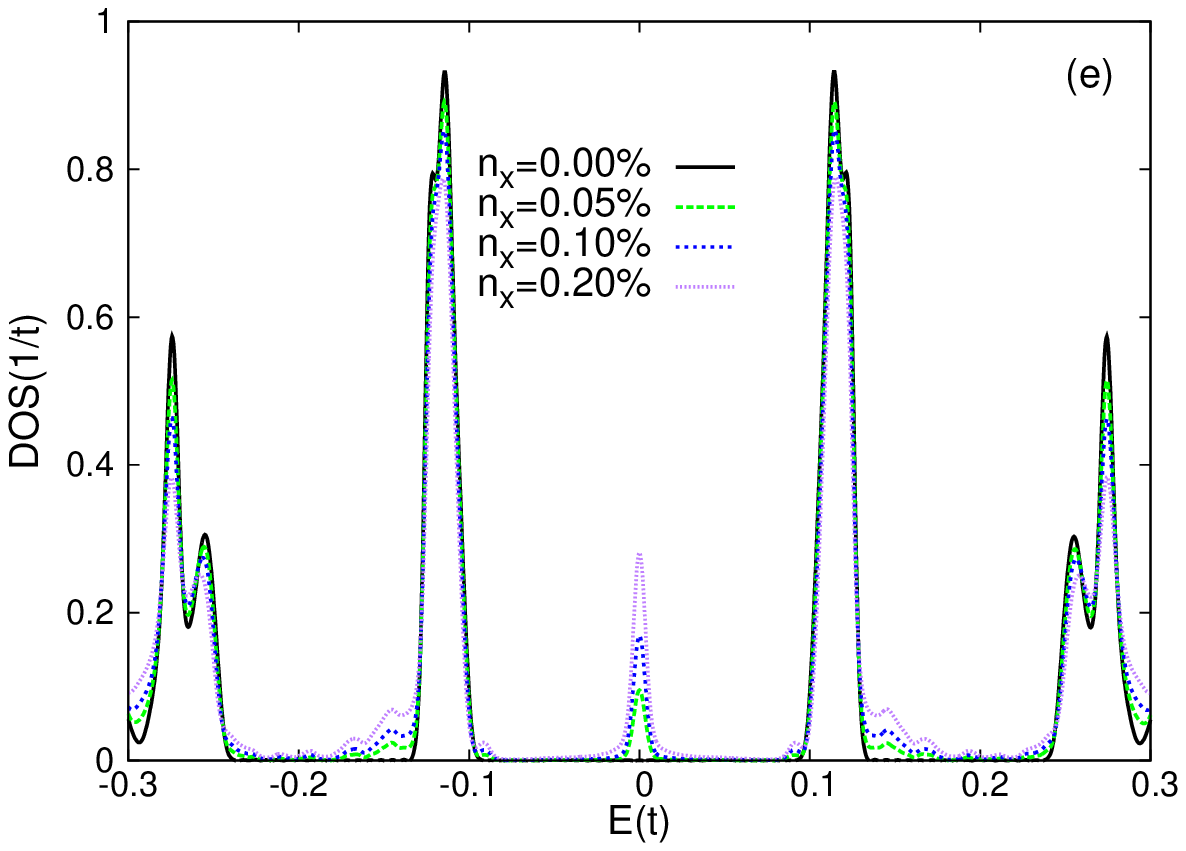}
\includegraphics[width=0.51\columnwidth]{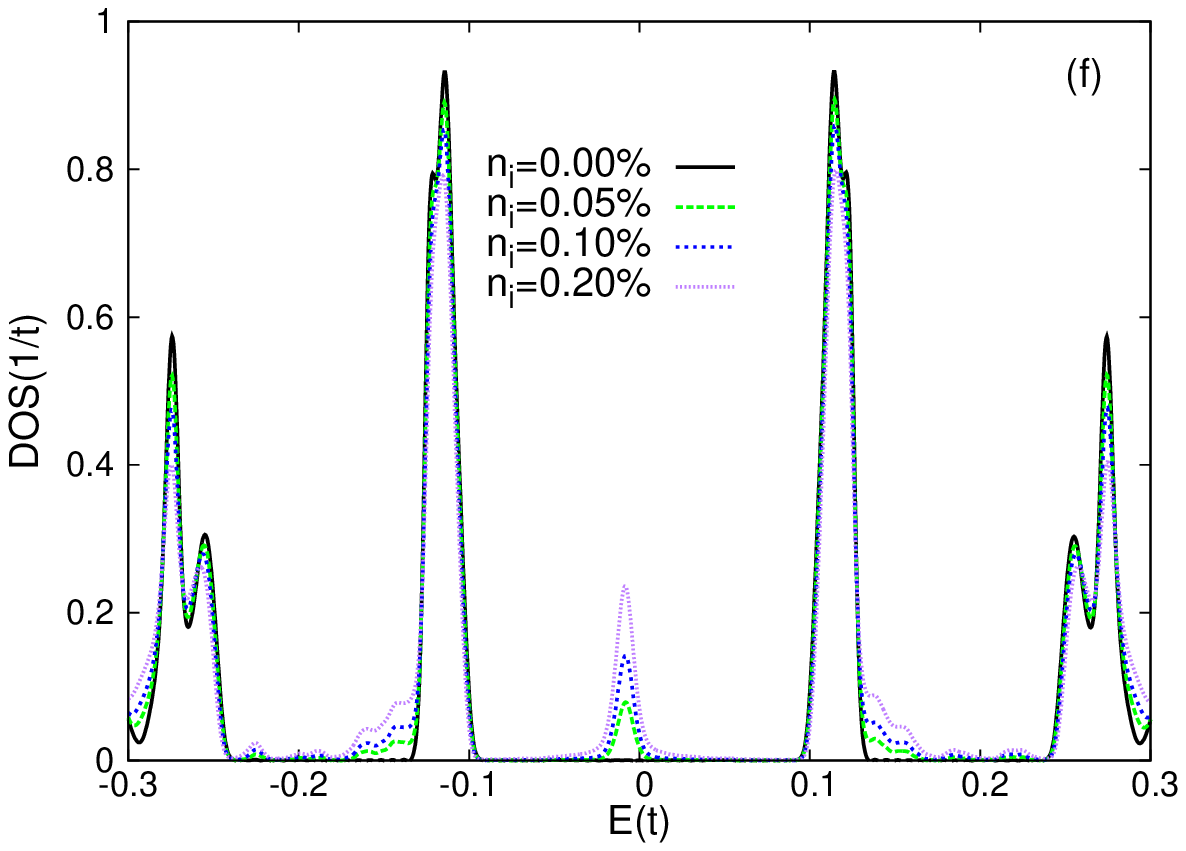}
}
\end{center}
\caption{(a) DOS of a $\{10,6\}$ GAL (black lines) and of SLG (red lines).
(b) Same as (a) but zoomed in low energies. (c) Random shift of the center
of the antidots within the range $(x\pm l_{C},y\pm l_{C})$. (d) Random
variation of the antidot radius within the range $[R-r_{R},R+r_{R}]$. (e)
Random vacancies, for four concentrations. (f) Random hydgrogen adsorbants,
for four coverages. The dotted vertical lines in panel (b) correspond to the
two different Fermi energies considered in this work: undoped GAL with $%
\protect\mu=0$ and doped GAL with $\protect\mu=0.115t$. }
\label{Fig:DOS}
\end{figure}

In this section we present the results for the polarization and dielectric
function of GALs, analyzing their main differences with respect to SLG. For
completeness, we summarize the salient results for the effect of the antidot
lattice on the DOS (both pristine and disordered systems), as discussed in
our earlier papers.\cite{PP08,YRJK13} The DOS of a $\{10,6\}$ GAL follows
from Eq. (\ref{Eq:DOS}), and the results are shown in Fig. \ref{Fig:DOS}.
The lattice of antidots splits the broad $\pi $ and $\pi ^{\ast }$ bands of
SLG into a number of gapped, narrow and flat bands, modifying the DOS of SLG
[given by the red line in Fig. \ref{Fig:DOS}(a)-(b)] into a set of peaks
associated to the Van Hove singularities of the subbands of the antidot
lattice [black lines in Fig. \ref{Fig:DOS} (a)-(b)].
Figs. \ref{Fig:DOS}(c)-(d) show the modification in the DOS of GALs due to
\textit{geometrical} disorder, associated to irregularities in the antidot
lattice, such as changes in the center-to-center distance of the etched
holes, or to variations in the size of the holes. In general, the gaps are
rather robust against geometrical disorder, and only after a large deviation
of the GAL array from the perfect periodicity, the gaps close. We further
show the effects of resonant impurities, such as vacancies or adatoms, in
the DOS of GALs [Fig. \ref{Fig:DOS}(e)-(f)]. This kind of defects leave the
structure of the DOS practically unchanged, apart from the creation of a
midgap band associated to localized states around the impurities, which
leads to the peak at $E\approx 0$ in the DOS.

\subsection{Electron-hole continuum and collective modes}

\begin{figure*}[t]
\begin{center}
\mbox{
\includegraphics[width=0.85\columnwidth]{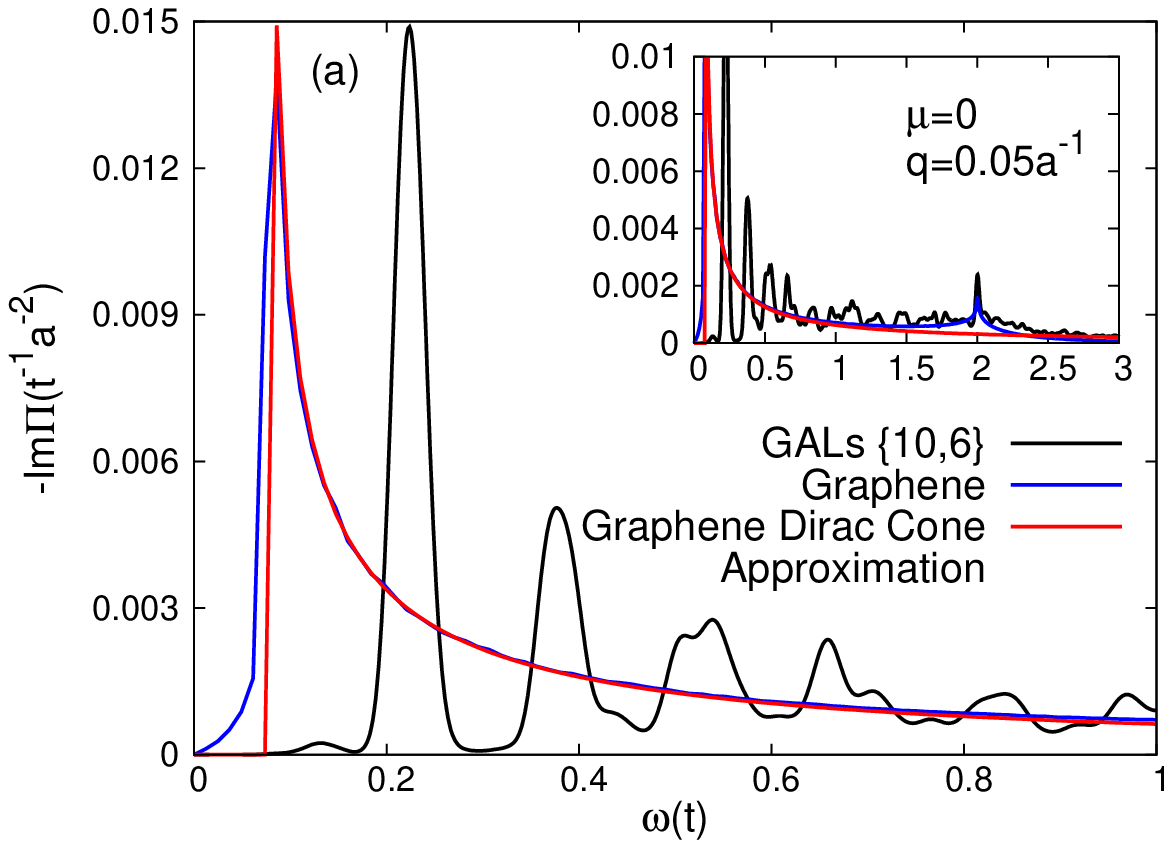}
\includegraphics[width=0.85\columnwidth]{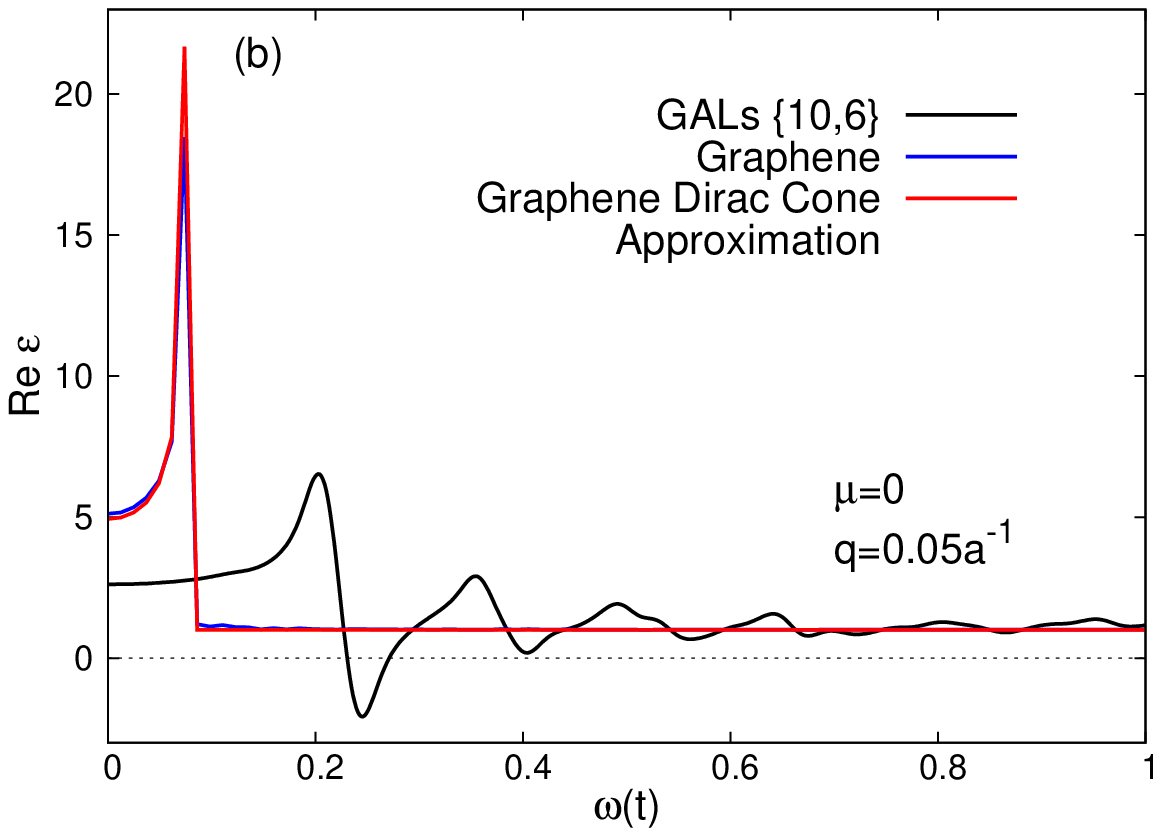}
}
\mbox{
\includegraphics[width=0.85\columnwidth]{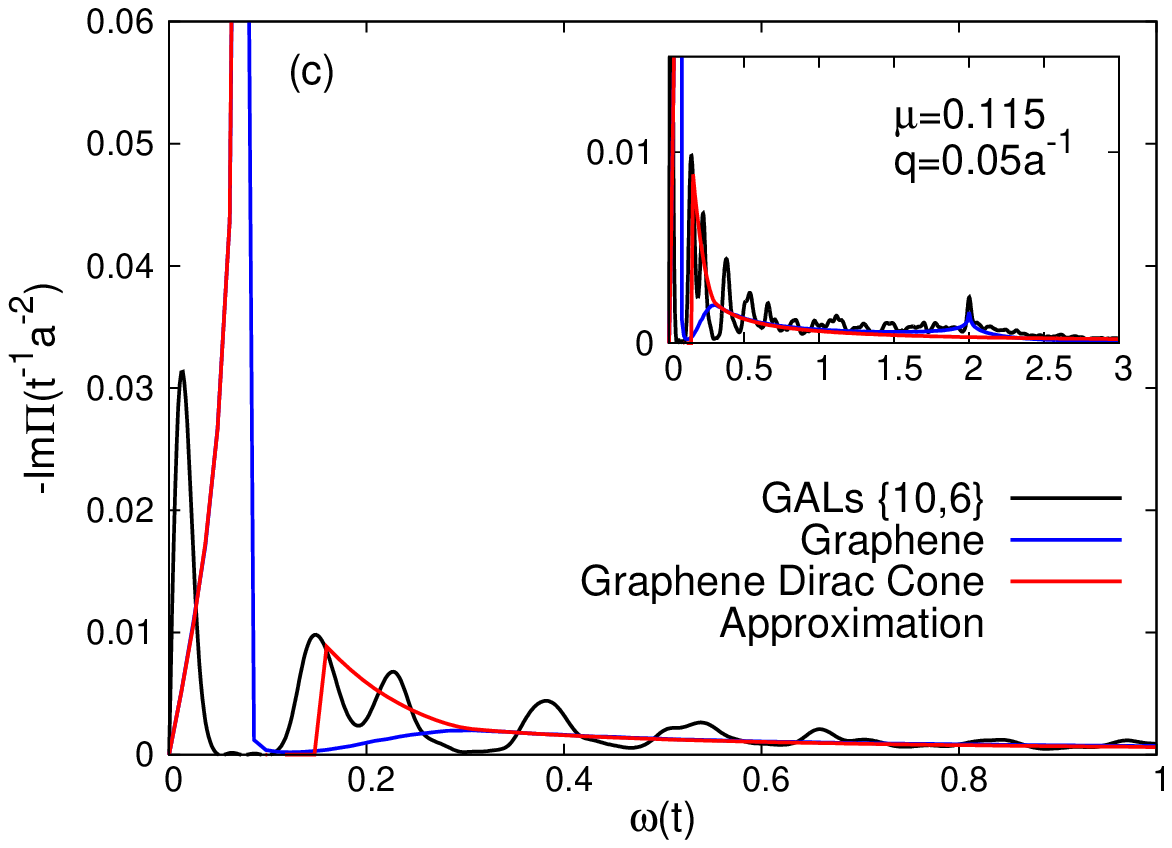}
\includegraphics[width=0.85\columnwidth]{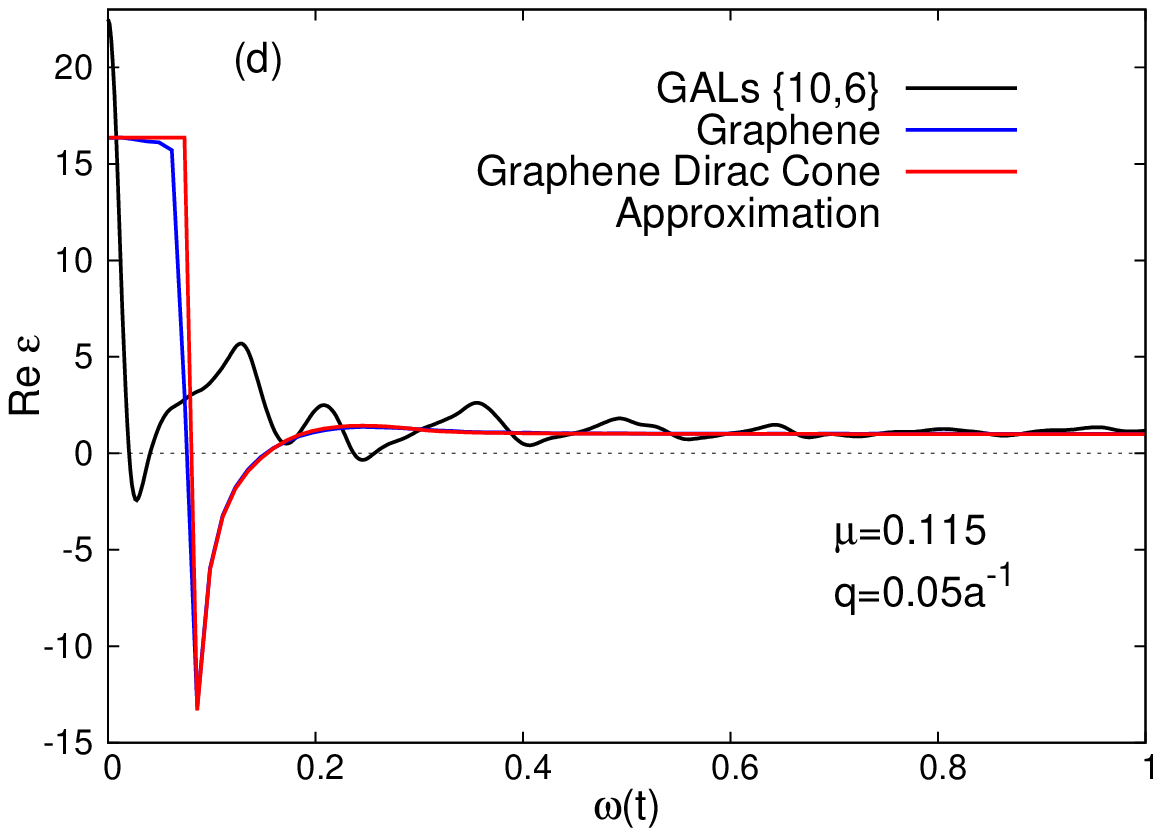}
}
\end{center}
\caption{Comparison between SLG and GALs. Polarization function [(a) and
(c)] and dielectric function [(b) and (d)] for a $\{10,6\}$ GAL. Black lines
correspond to GAL, whereas the blue lines correspond to SLG
within the $\protect\pi$-band tight-binding model, and the red lines
correspond to SLG using the Dirac cone approximation. Panels (a)
and (b) are for undoped samples ($\protect\mu=0$), and panels (c) and (d)
are for doped samples with $\protect\mu=0.115t$. The insets in (a) and (c)
give the spectra in a broader range of energies. }
\label{Fig:PiClean}
\end{figure*}

The structure of the DOS discussed above will be useful to understand the
polarization and dielectric functions of GAL. We first consider the case of
a clean GAL. Our results for $\Pi (q,\omega )$ and $\varepsilon (q,\omega )$
as a function of frequency $\omega $, are shown in Fig. \ref{Fig:PiClean}.
In the undoped regime [Figs. \ref{Fig:PiClean}(a)-(b)] we obtain an $\mathrm{%
Im}\Pi $ which consists of a series of peaks (the black line in Fig. \ref%
{Fig:PiClean}(a)), which indicate the electron-hole continuum in GALs,
defined as the region of the energy/momentum plane where particle-hole
excitations are possible. For comparison, we also show the polarization
function for undoped SLG from a $\pi $-band tight-binding model\cite%
{YRK11} (blue line) and within the Dirac cone approximation (red
line). In both cases, undoped graphene and
undoped GAL, only inter-band transitions are possible. However, there are
strong differences in the two spectra. On the one hand, the inter-band
electron-hole excitation spectrum in SLG is a continuum which (near the
Dirac point) corresponds to the region in the $\omega -q$ plane in which $%
\omega >v_{F}q$, where $v_{F}$ is the Fermi velocity of graphene near the
Dirac point. In fact, $\mathrm{Im}~\Pi (q,\omega >v_{F}q)$ is a monotonic
function for a broad range of energies, Fig. \ref{Fig:PiClean}(a). At high energies, of the
order of $\omega \approx 2t\sim 5$~eV, the spectral function shows a peak
which is associated to inter-band transitions between states of the Van Hove
singularities of the valence and of the conduction bands [see inset of Fig. %
\ref{Fig:PiClean}(a)]. Notice that this effect is not captured by the Dirac
cone approximation (red line), which is valid only for transitions in the
vicinity of the Dirac point. The electron-hole excitation spectrum is
qualitatively different for GALs, the black line of Fig. \ref{Fig:PiClean}%
(a). Most saliently, there are regions of zero spectral weight in $\mathrm{Im%
}~\Pi (q,\omega )$, which are associated to the inter-band gaps opened in
the GAL band structure, as compared to the continuum spectrum associated to
the broad $\pi $ and $\pi ^{\ast }$ bands in graphene [see e. g. Figs. \ref%
{Fig:DOS}(a) and (b) for the corresponding DOS].

Also the dielectric function of undoped GALs and SLG are qualitatively
different, as shown in Fig. \ref{Fig:PiClean}(b). The oscillatory DOS of the
GAL leads to a dielectric function with an oscillatory behavior, too. Second,
focusing on collective excitations, which are obtained from Eq. (\ref%
{Eq:Plasmons}), we observe that $\mathrm{Re}~\varepsilon$ vanishes for
undoped GALs \textit{at least once}, whereas there is no solution of Eq. (%
\ref{Eq:Plasmons}) for undoped graphene (the blue and red lines in Fig. \ref%
{Fig:PiClean}(b)). This is the well-known result about the absence of
plasmons in undoped SLG within the RPA.\cite{S86,WSSG06,HS07,RGF10} The
situation is rather different for GALs, whose peculiar band structure and
DOS lead to a collective mode which is associated to inter-band
electron-hole transitions between the valence and conduction bands, with
energies (for the $\{10,6\}$ structure) $E \approx \pm 0.12t$, as it is
shown in Fig. \ref{Fig:DOS}(b). However, this mode cannot be considered as a
fully coherent plasmon since its dispersion lies in a region of the spectrum
where $\mathrm{Im}~\Pi(\mathbf{q},\omega_{pl})\ne 0$. Therefore, such a mode
will be \textit{damped}, decaying into electron-hole pairs. It is interesting
to notice that this mode has a similar origin as the so-called $\pi$-plasmon
mode in graphene,\cite{YRK11} which is a \textit{damped} mode in
single-layer and multi-layer graphene originated from particle-hole
transitions between the Van Hove singularity of the valence band at $%
E\approx -t \sim -2.7$~eV and the Van Hove singularity of the conduction
band at $E\approx +t \sim +2.7$~eV, and which has been observed by EELS
experiments.\cite{EB08}

\begin{figure*}[t]
\begin{center}
\mbox{
\includegraphics[width=0.85\columnwidth]{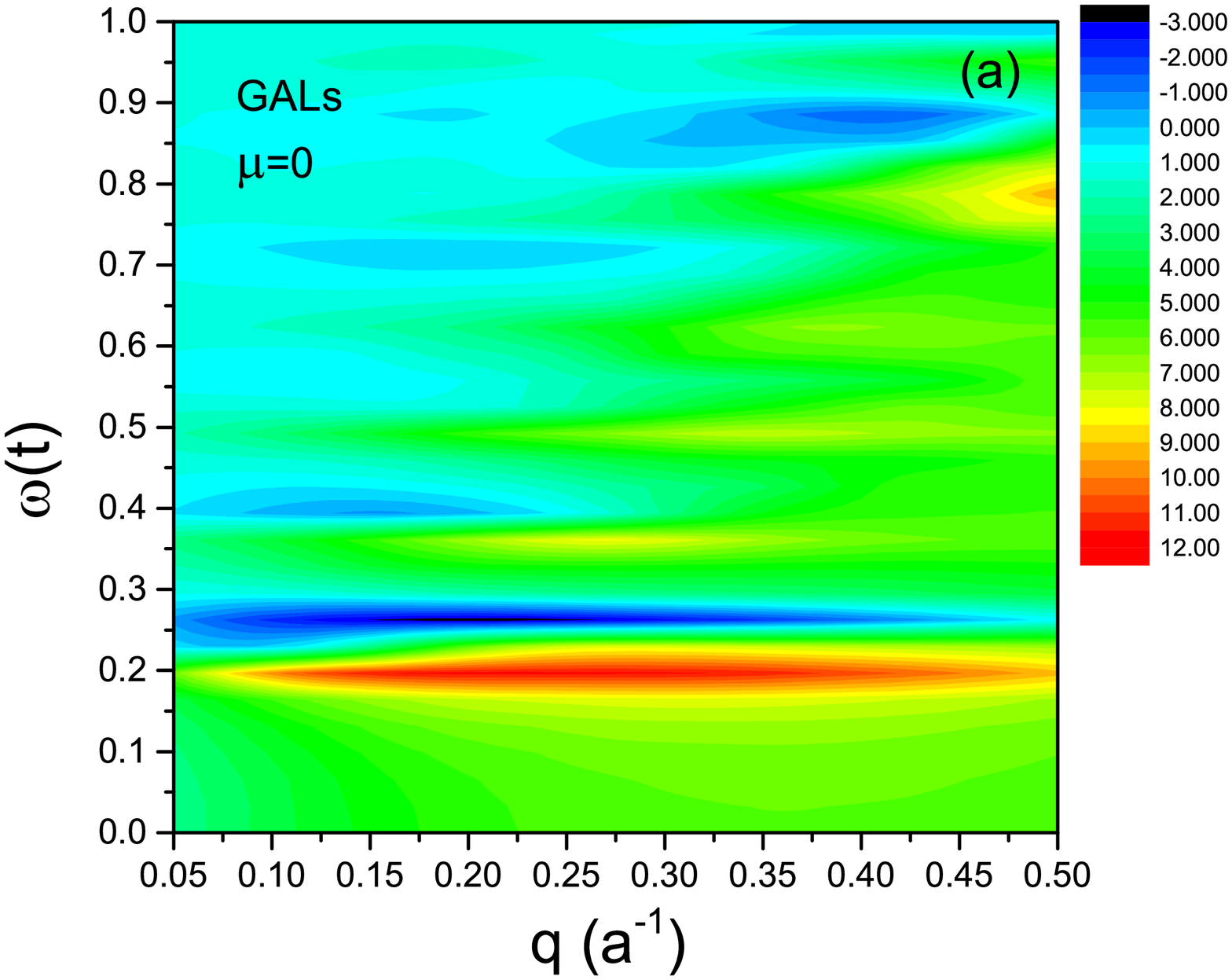}
\includegraphics[width=0.85\columnwidth]{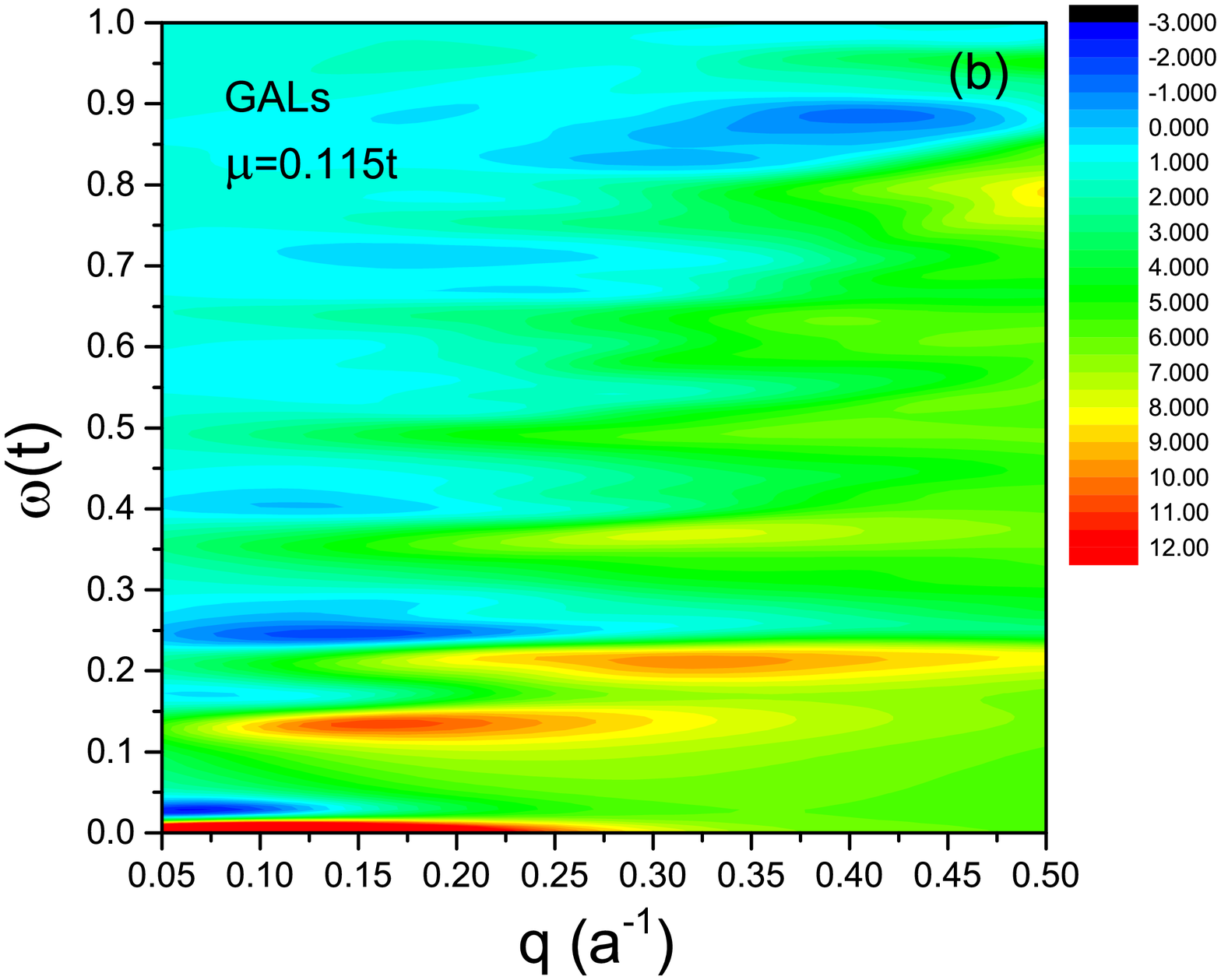}
}
\mbox{
\includegraphics[width=0.85\columnwidth]{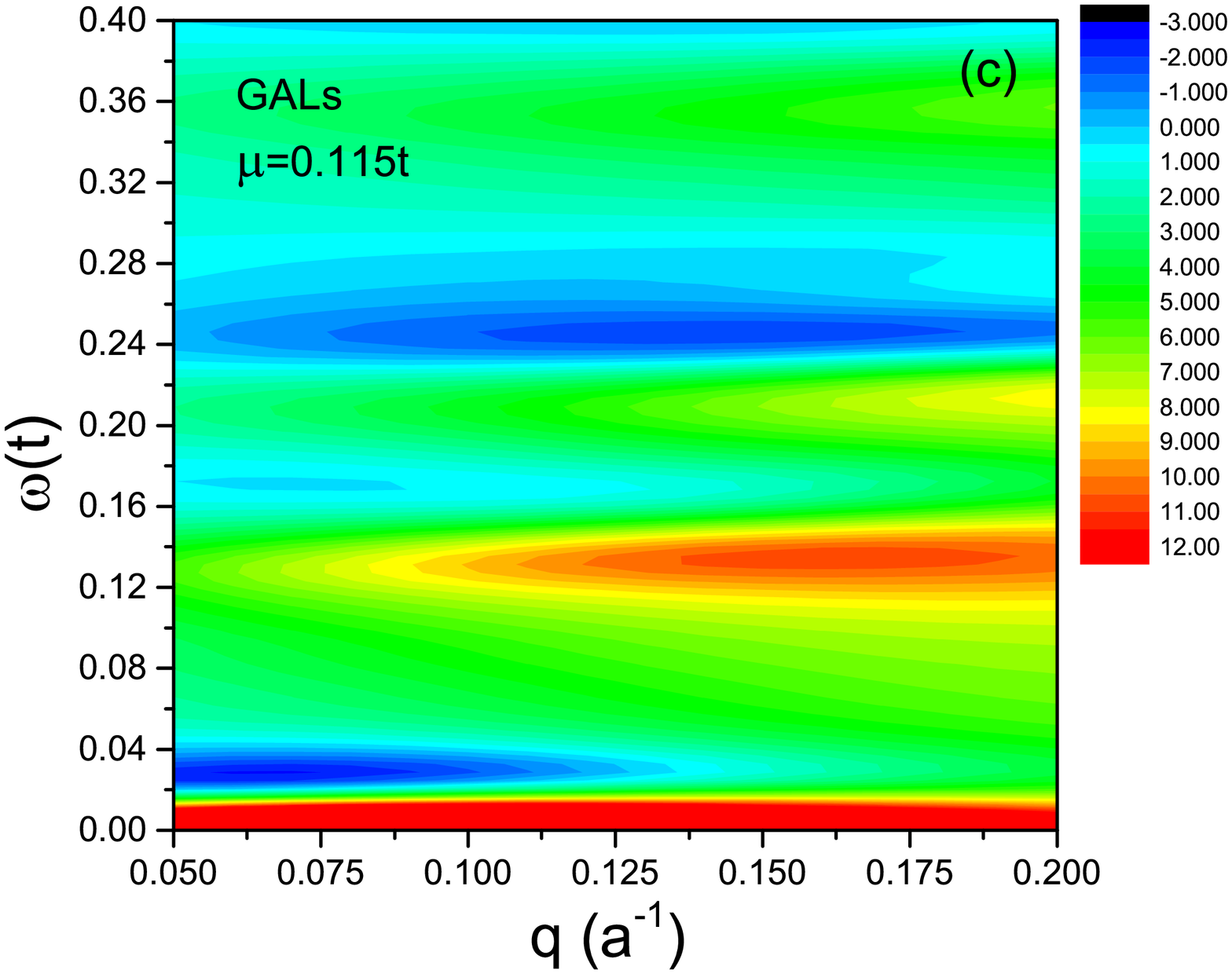}
\includegraphics[width=0.85\columnwidth]{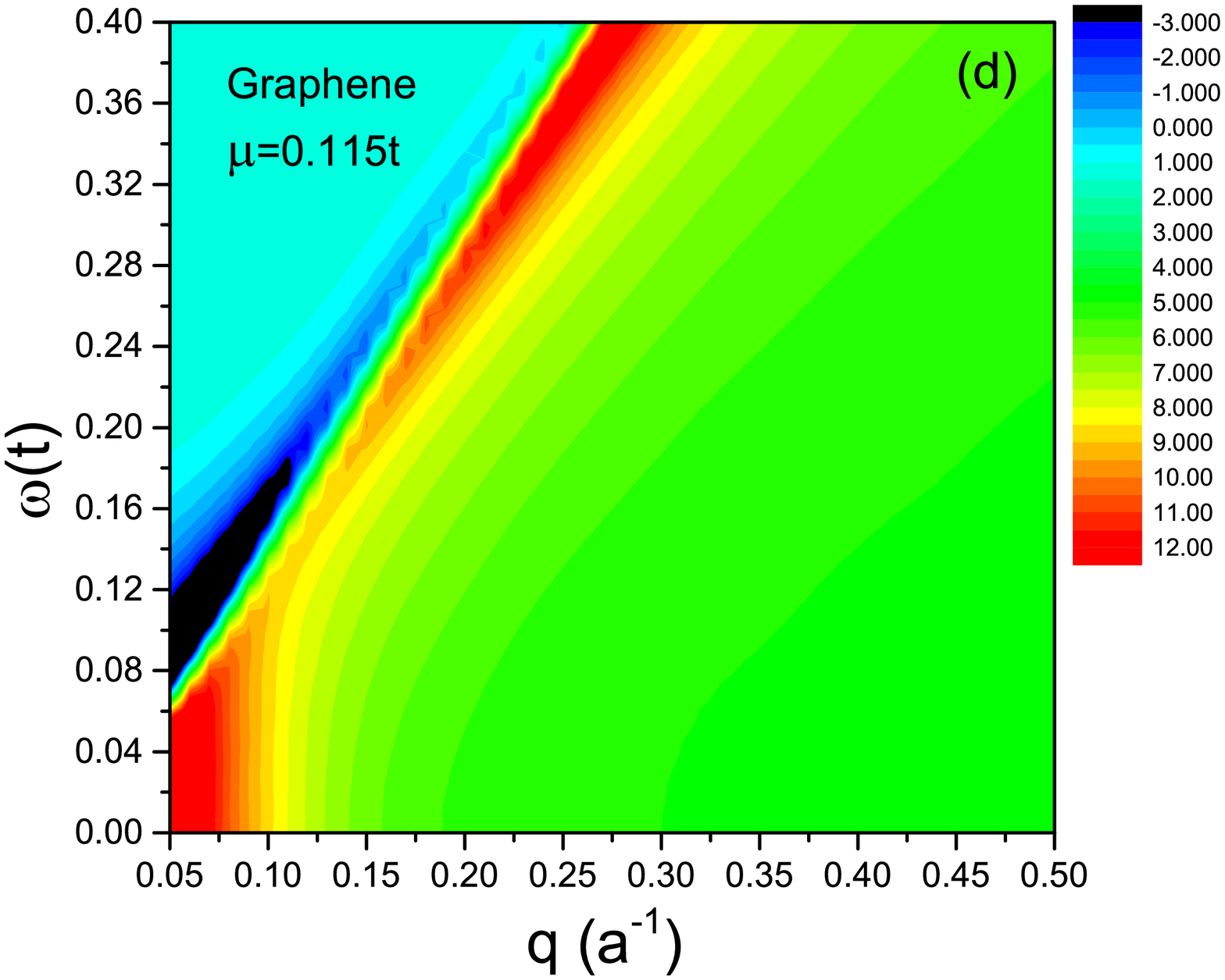}
}
\end{center}
\caption{Density plots of $\mathrm{Re}~\protect\varepsilon(q,\protect\omega)$%
. (a) Undoped ($\protect\mu=0$) GAL with $\{10,6\}$ periodicity. (b) Doped
GAL with $\protect\mu=0.115t$. (c) Same as (b) for a reduced region of the
spectrum corresponding to small wave-vectors and frequencies. (d) Doped SLG
with $\protect\mu=0.115t$. Notice the different scales of frequencies and
momenta in each plot. Notice that the plots start at $q=0.05/a$.\protect\cite%
{Foot_q} }
\label{Fig:DensityPlot}
\end{figure*}

\begin{figure*}[t]
\caption{ Density plot of $\mathrm{Re}~\protect\varepsilon(q,\protect\omega)
$ for four different types of disorder. (a) Random center; (b) Random
radius; (c) Random vacancies; and (d) Random hydrogen adsorbants. The
parameters characterizing the disorder are indicated in the figures.}
\label{Fig:DensityPlotDisorder}
\begin{center}
\mbox{
\includegraphics[width=0.85\columnwidth]{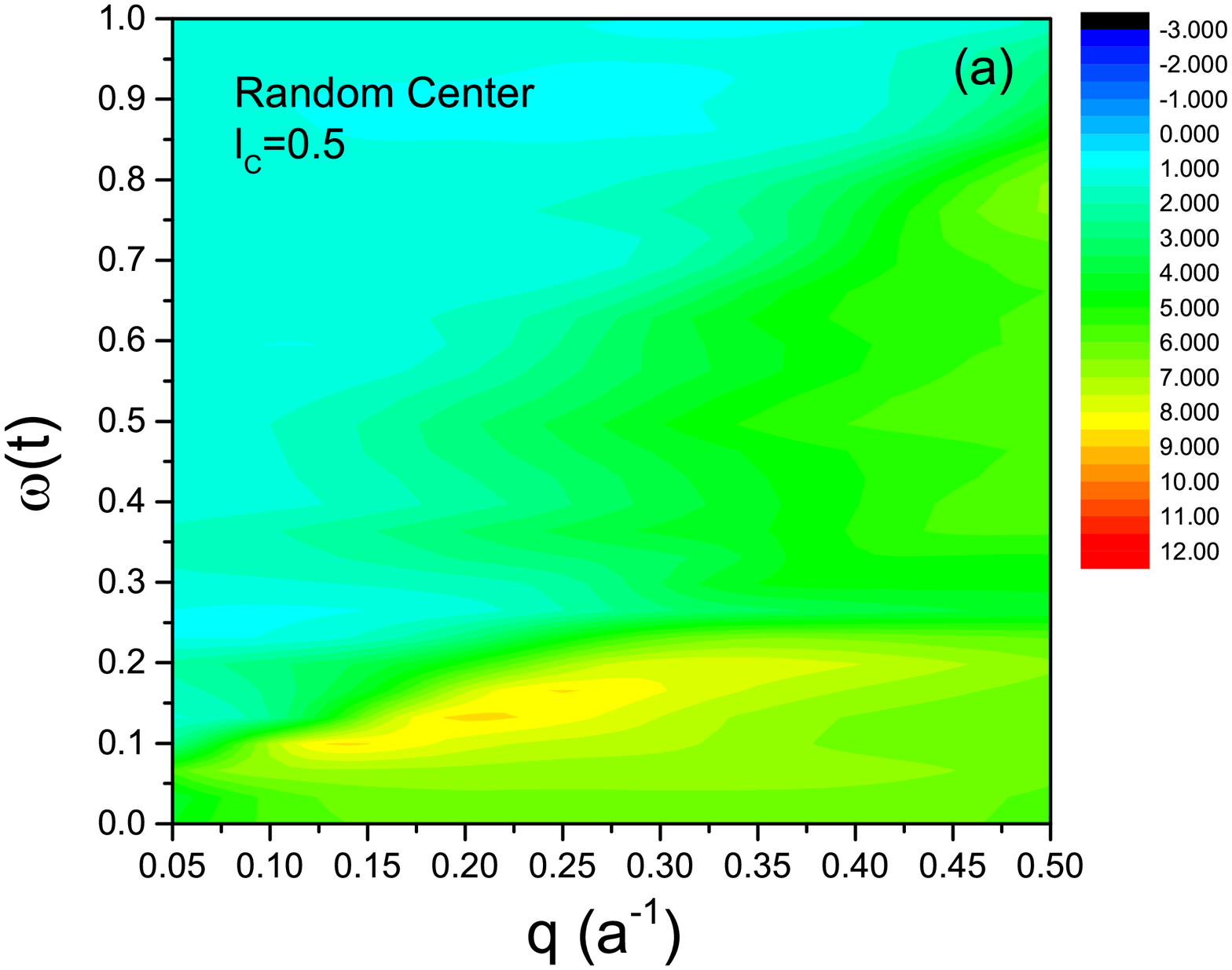}
\includegraphics[width=0.85\columnwidth]{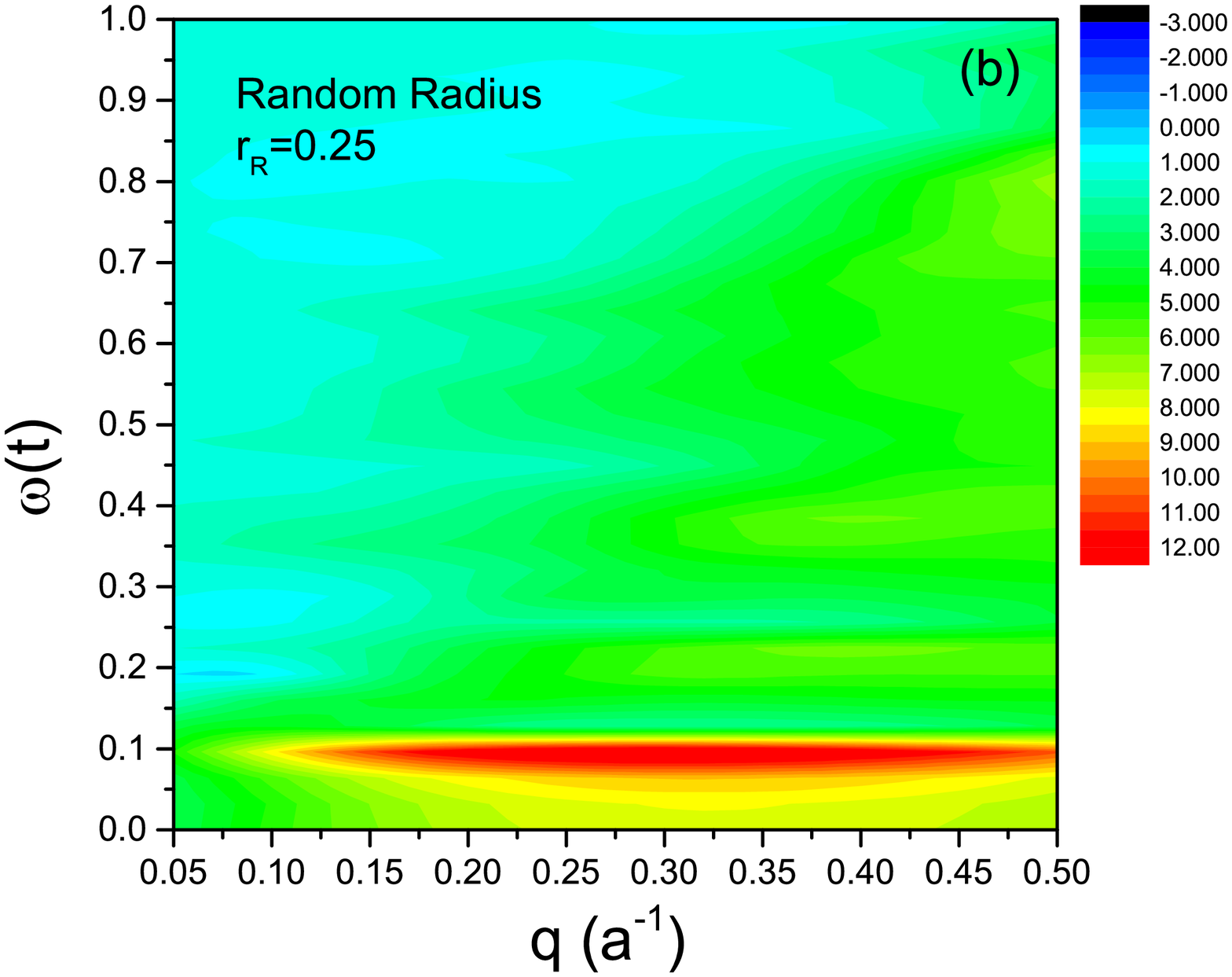}
}
\mbox{
\includegraphics[width=0.85\columnwidth]{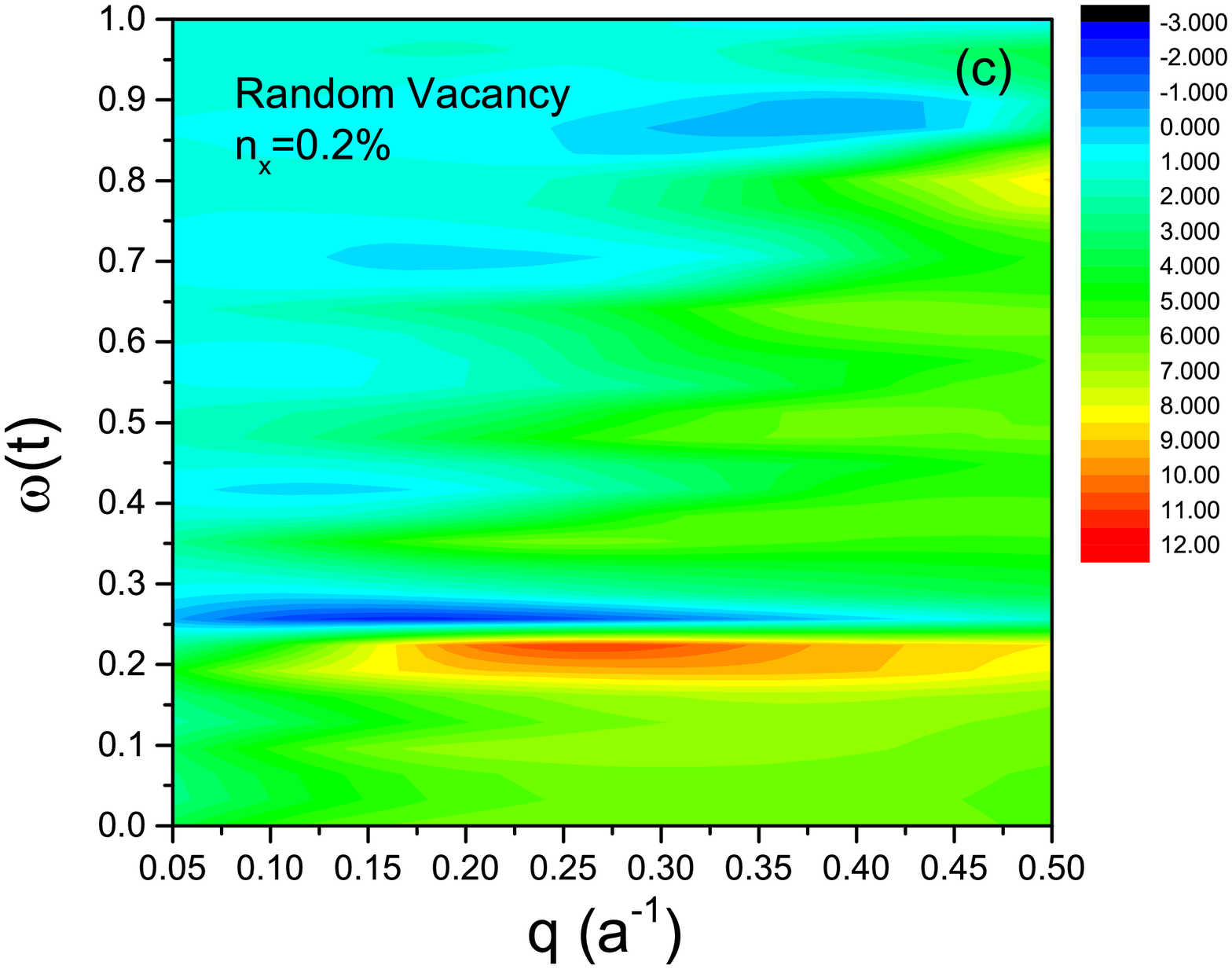}
\includegraphics[width=0.85\columnwidth]{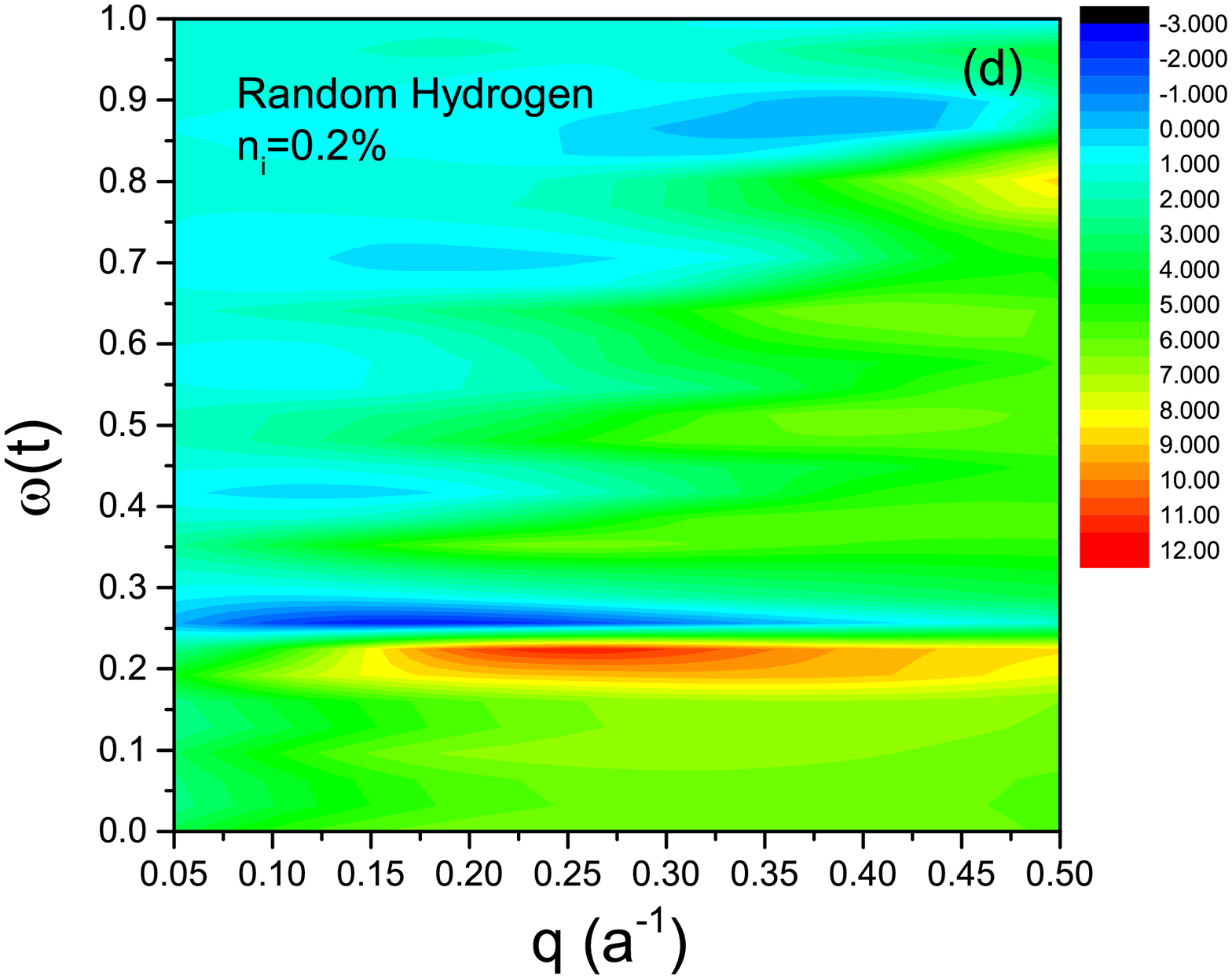}
}
\end{center}
\par
\end{figure*}

The results for doped GALs are shown by the black lines in Fig. \ref%
{Fig:PiClean}(c)-(d), which are compared to the results of SLG from a $\pi$%
-band tight-binding model (blue lines) and from the Dirac cone approximation
(red lines). The chemical potential $\mu\approx 0.115t$ is such that the
Fermi energy lies within the first peak in the DOS, as shown by the
corresponding dotted black line of Fig. \ref{Fig:DOS}(b). Here we define the
chemical potential as measured from the $E=0$ Dirac point energy in
graphene. In GAL, since it is a semiconductor, it is useful to define $%
\mu\approx\Delta/2 + E_F$, where $\Delta$ is the gap energy and $E_F$ is the
Fermi energy measured, as usual, from the bottom of the conduction band. One
first notices that $\mathrm{Im}~\Pi(q,\omega)$ has two different
contributions: one at low energies, which is due to intra-band excitations,
and which is very narrow due to the small bandwidth $W$ of the doped band in
GAL. This bandwidth $W$ imposes a limit for the energy of intra-band
transitions. The second contribution emerges (for $q=0$) at $\omega\approx
\Delta + 2 E_F$, and is due to inter-band excitations across the band gap.
This is the counterpart of the inter-band electron-hole continuum in SLG,
and it is the only contribution in the $\mu=0$ case, as discussed above.
Interestingly, doping leads to a new plasmon mode in the spectrum, as seen
by the zeros of $\mathrm{Re}~\varepsilon(q,\omega)$ in Fig. \ref{Fig:PiClean}%
(d). This solution of Eq. (\ref{Eq:Plasmons}) corresponds to the usual
plasmon mode which, in this case, is long-lived due to the fact that it
disperses (for a certain range of energies) in the region of the $(q,\omega)$ space
where $\mathrm{Im}~\Pi(q,\omega)=0$. In fact, such a mode is present also in
SLG, as seen by the zero-energy cut of the blue line in Fig. \ref%
{Fig:PiClean}(d). However, for a given wave-vector, the energy at which $%
\mathrm{Re}~\varepsilon(q,\omega)=0$ is different for GAL and for SLG, and
hence the dispersion of the plasmon is very different in the two cases. In
particular, from our results one can expect that the \textit{velocity} of
the mode (given by the slope of the band dispersion) will be lower in GAL
than in SLG. Considering that in 2D the dispersion relation of the plasmons
at low energies can be approximated by\cite{GV05}
\begin{equation}  \label{Eq:Plasmon}
\omega_{pl}(q)\approx \sqrt{\frac{2\pi n e^2 q}{m_b}}
\end{equation}
where $n$ is the carrier density and $m_b$ is the effective mass of the
band, our numerical result for the existence of \textit{slow}
plasmons in GALs is expected due to the flatness of their bands [see Fig. %
\ref{Fig:Bands}(b)], which typically implies large effective masses.
Furthermore, we notice that the theoretical studies for the band structure
of GALs show that they have a direct gap, located in the $\Gamma$ point of
the Brillouin zone.\cite{PP08} The minimum of the conduction band and the
maximum of the valence band can be well approximated, in leading order, by a
parabolic band dispersion, whose effective mass is simply related to the
DOS, $d(\mu)$, by the usual relation $m_b=2\pi\hbar^2d(\mu)/g_{\sigma}$,
where $g_{\sigma}=2$ is the spin degeneracy. Therefore, taking into account
that the nano-perforation of the graphene lattice allows to \textit{%
manipulate} the DOS of the system, the corresponding dispersion relation of
the plasmon mode will be specific and different for each pair $\{L,R\}$ that
characterizes a given GAL.

These features are more clearly seen in Fig. \ref{Fig:DensityPlot}, where we
show a density plot of $\mathrm{Re}~\varepsilon (q,\omega )$ in the $\omega
-q$ plane. Consider first the undoped $\mu =0$ case, Fig. \ref%
{Fig:DensityPlot}(a). In the absence of charge carriers, the GAL does not
exhibit any low energy plasmon mode, and the first line of zeros of the
dielectric function corresponds to the gapped plasmon which is associated,
as discussed above, to inter-band transitions between states of the bands
adjacent to the Fermi level $\mu =0$. The energy of the mode at $%
q\rightarrow 0$ coincides with the gap opened in the band structure, $\Delta
\approx 0.24t$ for the present case, and it is almost dispersionless.\cite%
{Foot_q}  The reason for this weak dispersion of the gapped mode is
the narrowness and flatness of the bands involved in the inter-band
transitions. Apart from this well defined mode with a gap $\Delta \approx
0.24t$, Fig. \ref{Fig:DensityPlot}(a) also shows several resonances at
higher energies. They are signatures of electron-hole transitions of higher
energies, associated to the bands which lead to the different peaks in the
DOS of Fig. \ref{Fig:DOS}.
Similarly as in SLG,\cite{EB08} this structure should be accessible by
means of EELS experiments, which could be
useful to determine the main gap, as well as
the characteristic energies of the other flat bands, which would lead to
further resonances in the EELS spectrum.

The situation is different for doped GALs, Fig. \ref{Fig:DensityPlot}(b).
Apart from the gapped mode (with $\Delta \approx 0.24t$) discussed above, we
observe a strong feature at low energies which corresponds to the gapless
classical plasmon, with dispersion relation given by Eq. (\ref{Eq:Plasmon}).
The low energy and small-$q$ region of the spectrum of doped GAL is shown
for clarity in Fig. \ref{Fig:DensityPlot}(c), where we observe the existence
of a collective mode in a $\omega -q$ region of the spectrum which, for the
corresponding undoped case of Fig. \ref{Fig:DensityPlot}(a), is entirely
featureless. This collective mode is the counterpart of the well known
plasmon mode characteristic of a 2DEG\cite{stern67,AFS82} or doped SLG.\cite%
{S86,WSSG06,HS07,RGF10} For comparison, we show a density plot of $\mathrm{Re%
}~\varepsilon (q,\omega )$ for doped SLG in Fig. \ref{Fig:DensityPlot}(d),
where one can clearly observe the \textit{classical} plasmon mode with $%
\omega (q)\propto \sqrt{q}$.\cite{Foot_q} By comparing Fig. \ref%
{Fig:DensityPlot}(d) to Fig. \ref{Fig:DensityPlot}(c), we observe that the
slope of the plasmon mode in SLG is much higher than the corresponding slope
of the low energy plasmon in doped GAL. The qualitatively different band
structures of SLG and GAL explain the different behaviors of the plasmon
mode in the two systems. Furthermore, notice that the $\omega (q)\sim \sqrt{q%
}$ plasmon is the only active mode in doped SLG [Fig. \ref{Fig:DensityPlot}%
(d)], which does not present additional \textit{optical} resonances at
higher energies, as the ones present in the spectrum of GALs [Fig. \ref%
{Fig:DensityPlot}(b)].

\subsection{Effect of disorder}

We next address the effects due to disorder. We have considered two main
sources of disorder: \textit{geometrical} disorder, which accounts for
deviations of the GAL from the perfect periodicity, and \textit{resonant}
impurities, which can be associated to vacancies in the graphene lattice, or
to adatoms which can be adsorbed by the surface. The results are shown in
Fig. \ref{Fig:DensityPlotDisorder}. The modification of the band structure
and DOS due to geometrical disorder, as shown by Fig. \ref{Fig:DOS}(c)-(d),
leads to a dramatic effect on the structure of the spectrum, as can be seen
by comparing Figs. \ref{Fig:DensityPlotDisorder}(a)-(b) to Fig. \ref%
{Fig:DensityPlot}(a). Changes in the center-to-center separation of the
antidots leads to a blurring of the resonance peaks in the spectrum, as it
is seen in Fig. \ref{Fig:DensityPlotDisorder}(a), where the feature
associated to the gapped plasmon with energy $\Delta\approx 0.24t$ has
practically disappeared. In fact, no zeroes are present in the dielectric function after this kind of disorder is considered, what means that the plasmon mode has disappeared. 

\begin{figure*}[t]
\begin{center}
\mbox{
\includegraphics[width=0.7\columnwidth]{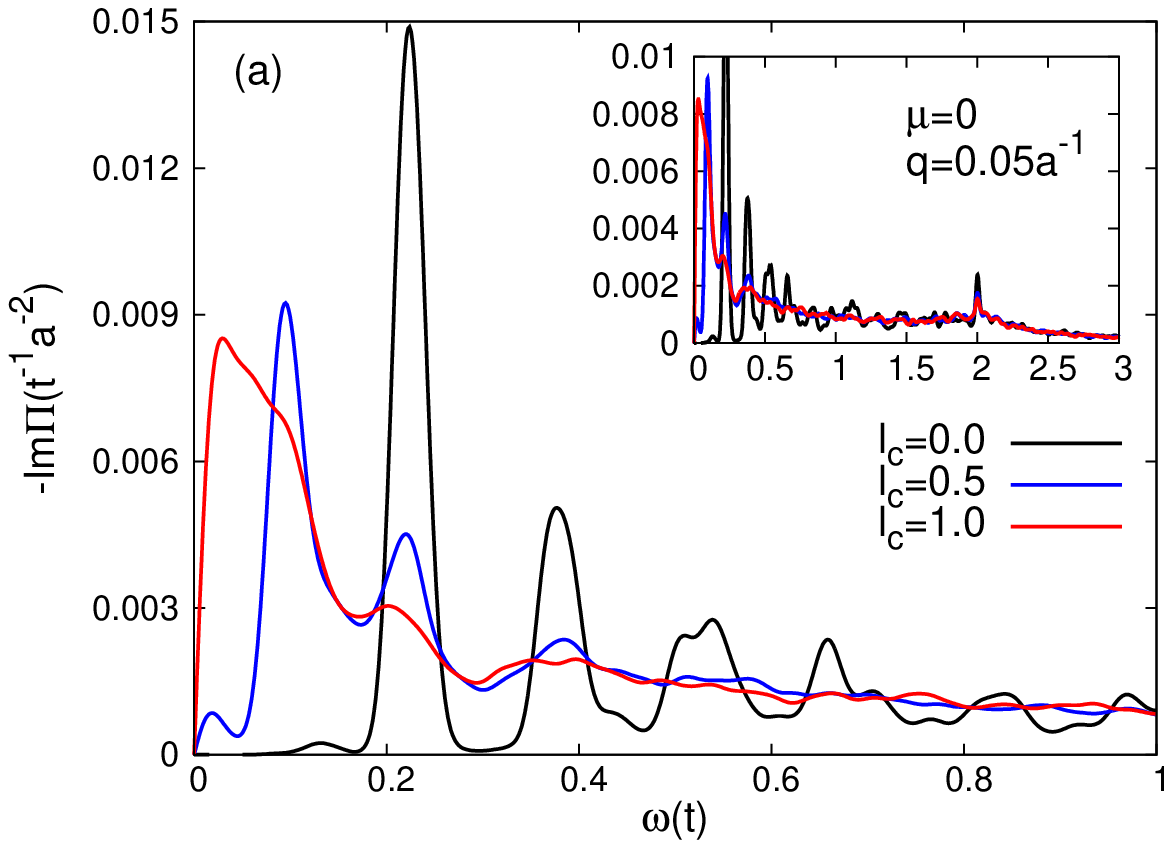}
\includegraphics[width=0.7\columnwidth]{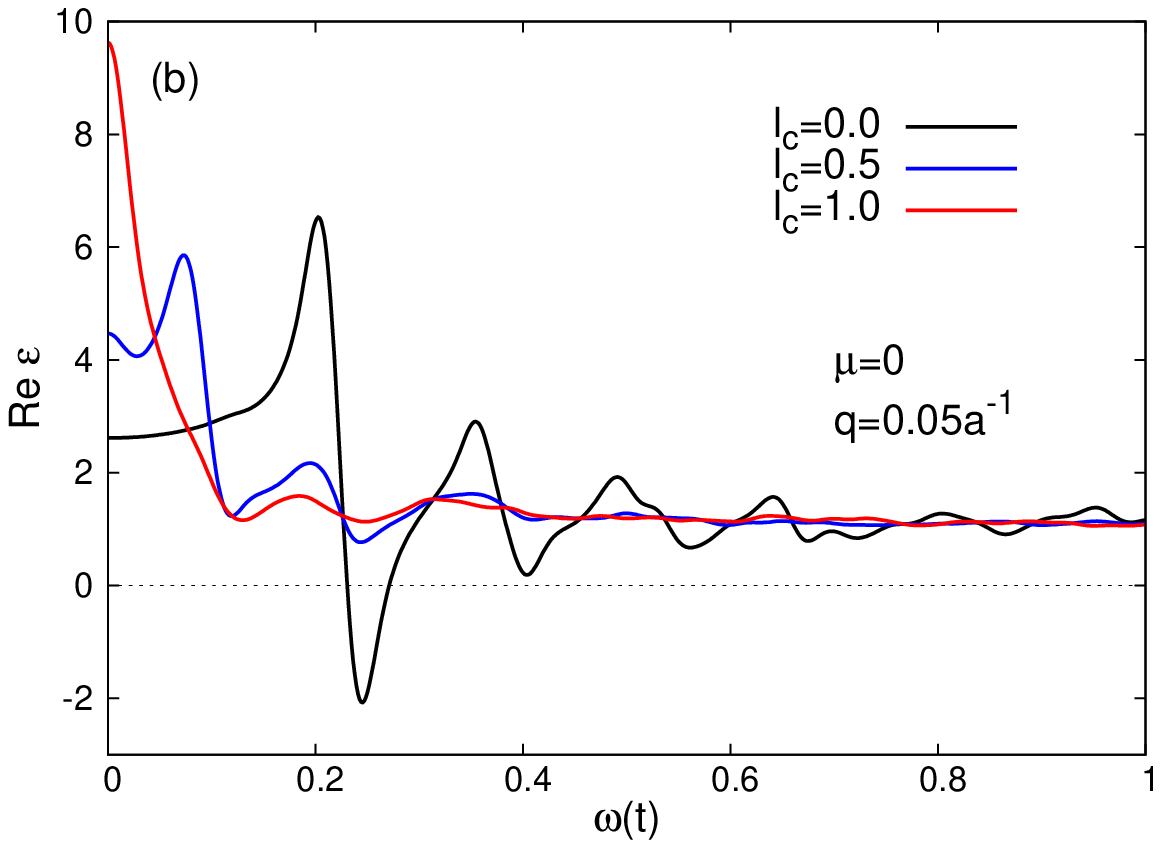}
}
\mbox{
\includegraphics[width=0.7\columnwidth]{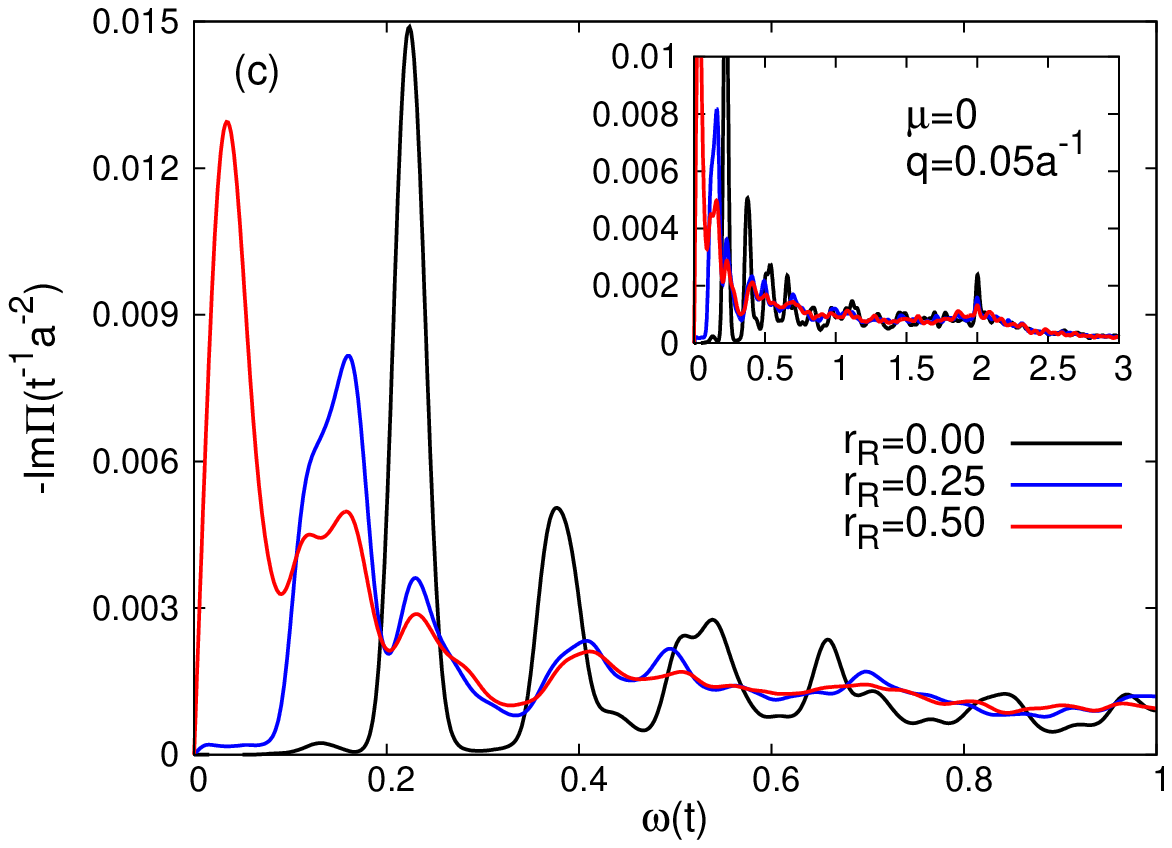}
\includegraphics[width=0.7\columnwidth]{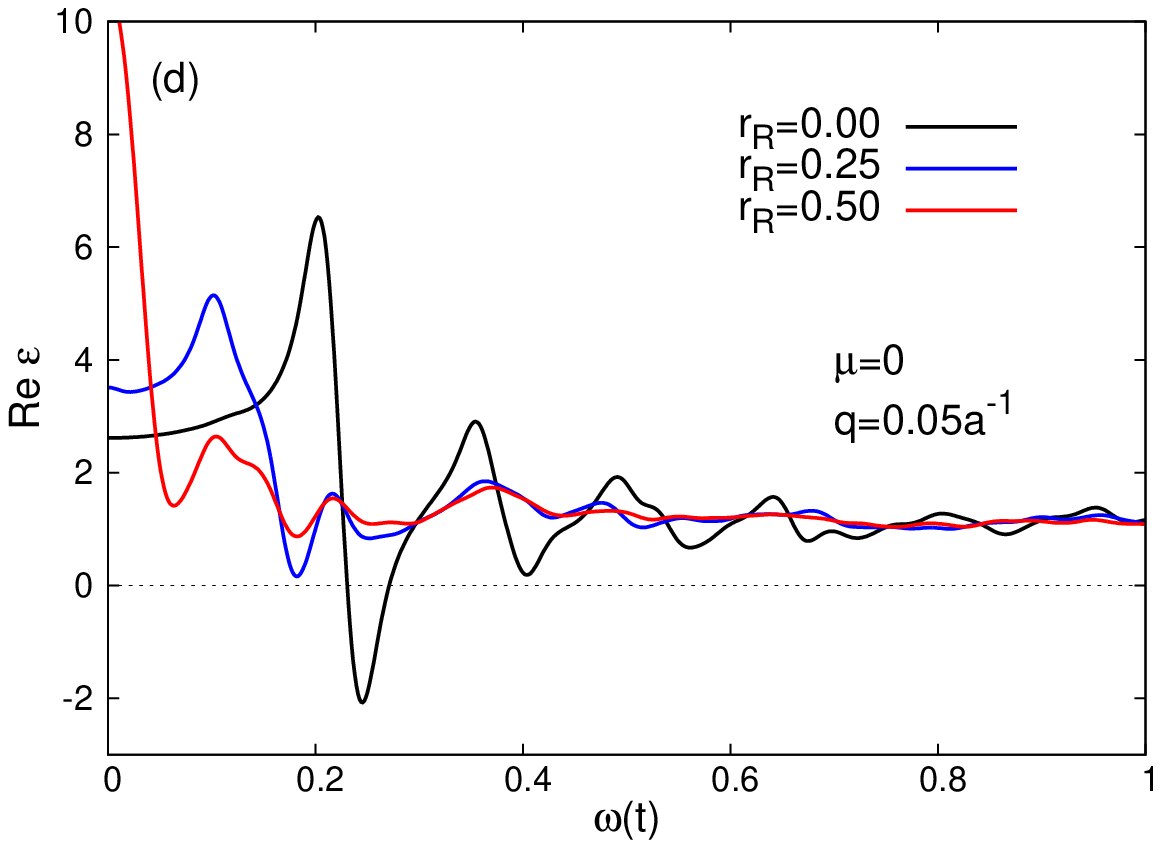}
}
\end{center}
\caption{Same as Fig. \protect\ref{Fig:PiClean} but in the presence of
\textit{geometrical} disorder. Panels (a) and (b) corresponds to $\mathrm{Im}%
~\Pi(q,\protect\omega)$ and $\mathrm{Re}~\protect\varepsilon(q,\protect\omega%
)$, respectively, for a GAL in which the center of the holes is shifted
randomly with respect to the original position in the perfect periodic
array, within the range $(x\pm l_{C},y\pm l_{C})$. The different colors
correspond to different values of $l_C$ (in units of $a$), as denoted in the
inset of the figures. Panels (c) and (d) corresponds to a GAL where the
radius of the holes is randomly shrunk or enlarged within the range $%
[R-r_{R},R+r_{R}]$. Different colors correspond to different values of $r_R$
(in units of $a$). The insets in panels (a) and (c) show the polarization in
a broader range of energies.}
\label{Fig:PiGeometrical}
\end{figure*}

In Fig. \ref{Fig:PiGeometrical}
we show results for the polarization and dielectric function of GAL,
obtained from Eqs. (\ref{Eq:Kubo}) and (\ref{Eq:Epsilon}), in the presence
of geometrical disorder  for the wave-vector $q=0.05a^{-1}$. The presence of geometrical disorder can lead to the disappearance of the plasmon mode. This can be seen in Fig. \ref{Fig:PiGeometrical}(b) and (d), where there is an absence of zeros for the dielectric function of disordered GALs (blue and red lines), whereas ${\rm Re}~\varepsilon$ for clean GAL clearly shows a solution for the plasmon equation (\ref{Eq:Plasmons}). Furthermore, the well separated bands of clean GAL leads to a discretization of the electron-hole continuum at low energies. This can be seen by looking at the black line in Fig. \ref{Fig:PiGeometrical}(a) and (b), which shows an absence of spectral weight between the well defined peaks of ${\rm Im}~\Pi$ at low energies. However, the presence of geometrical disorder strongly modifies the band structure, and as a consequence, there is a transfer of spectral weight to the gapped regions of the spectrum.

\begin{figure*}[t]
\begin{center}
\mbox{
\includegraphics[width=0.7\columnwidth]{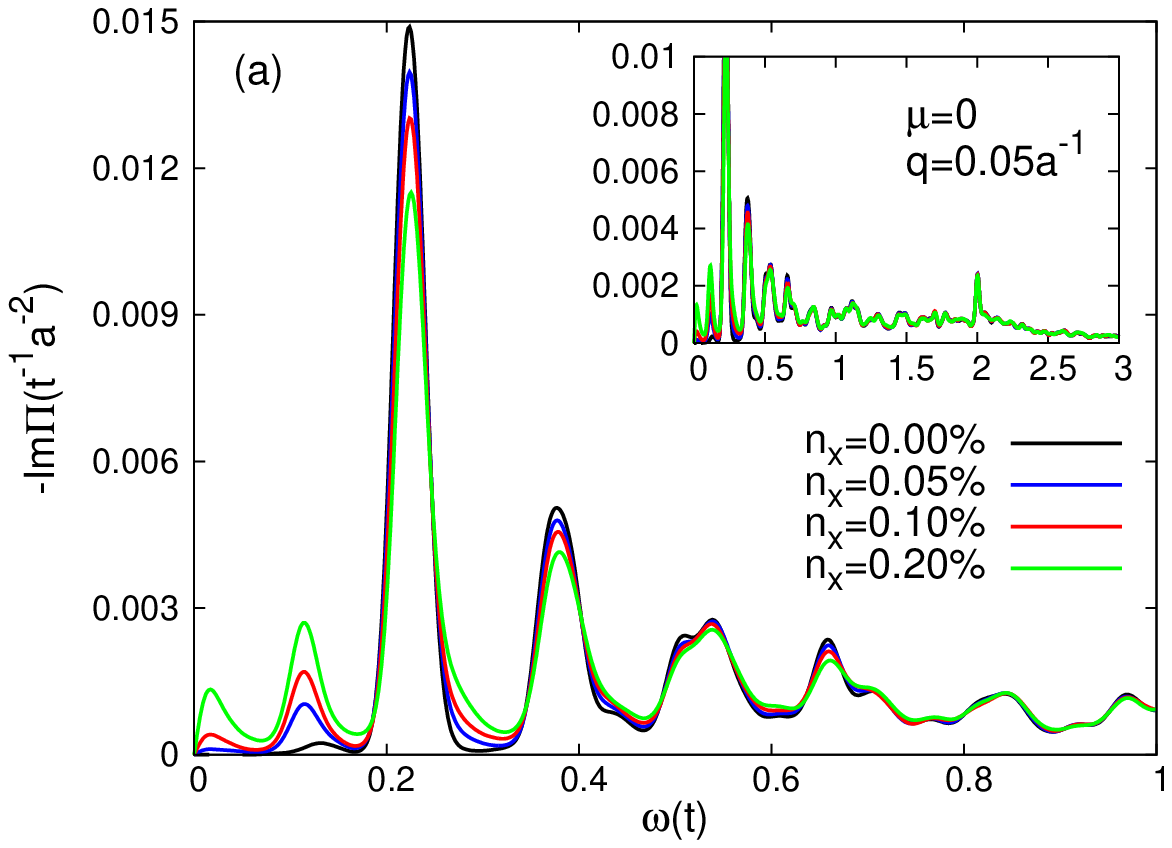}
\includegraphics[width=0.7\columnwidth]{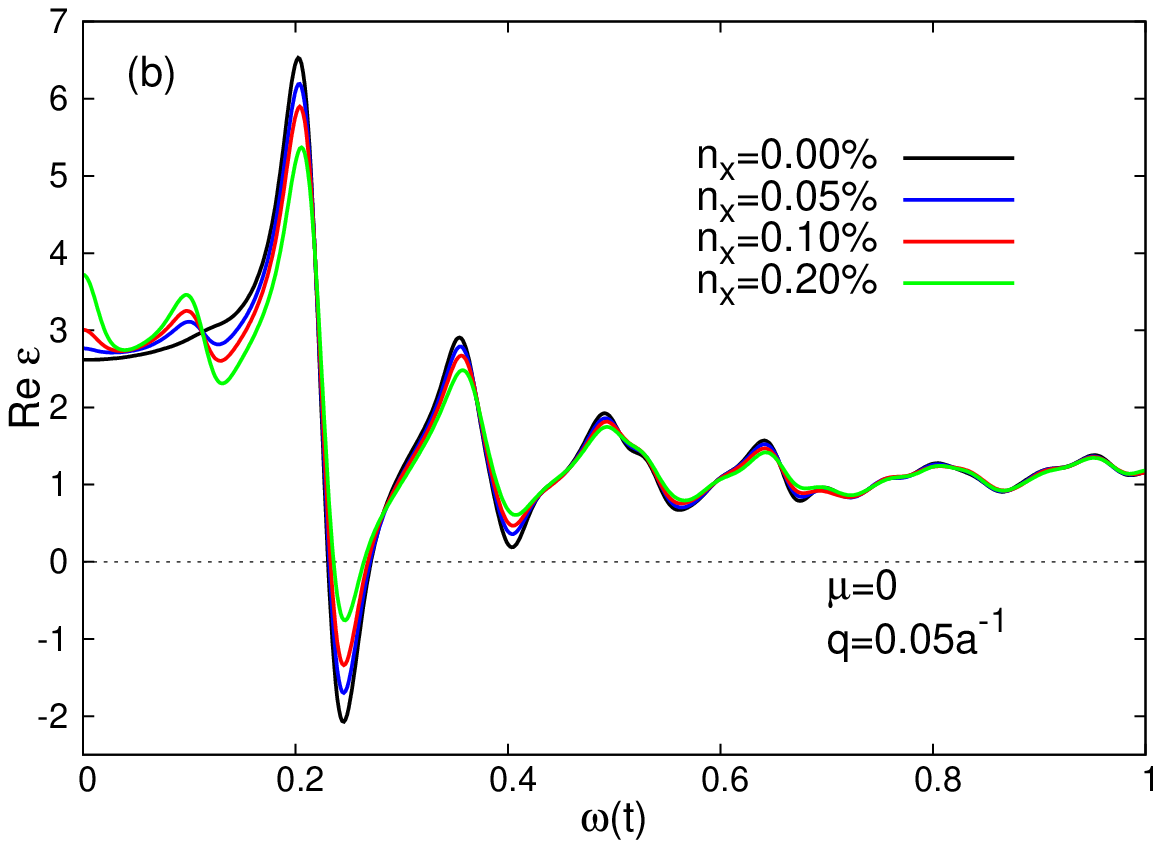}
}
\mbox{
\includegraphics[width=0.7\columnwidth]{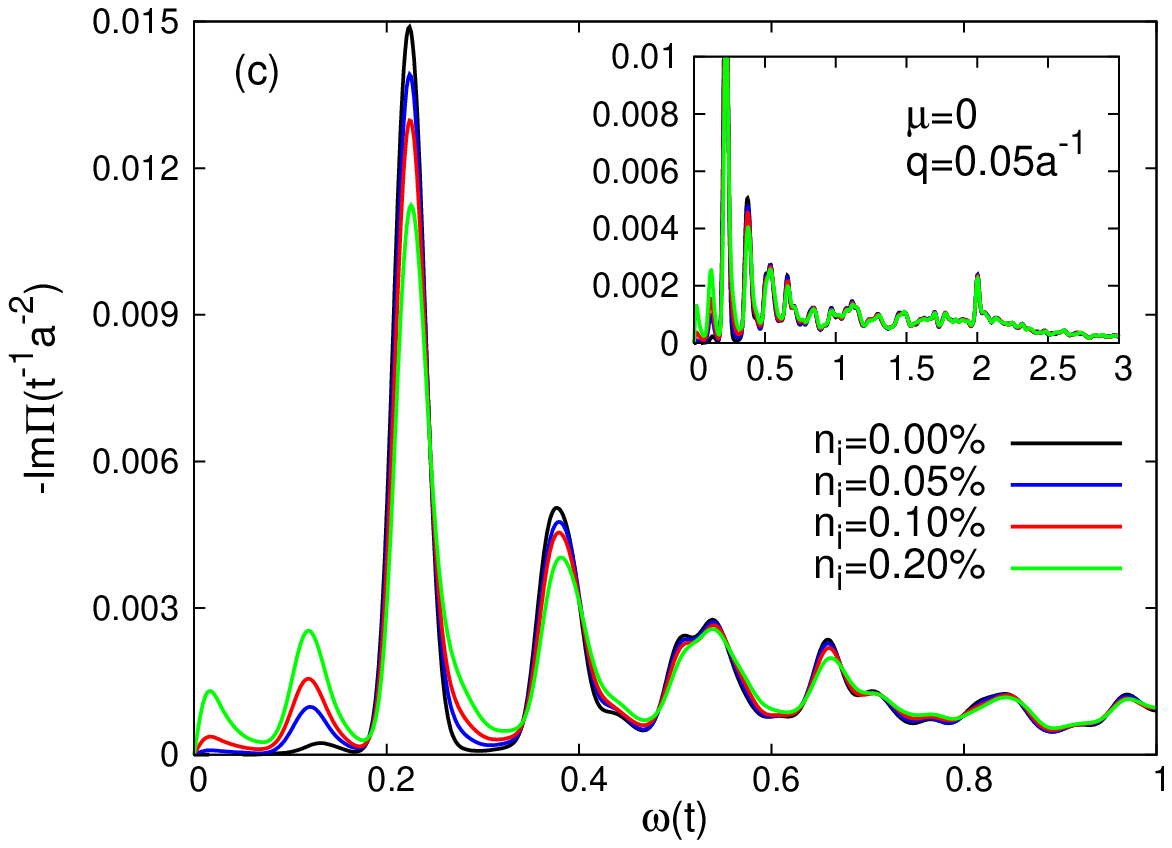}
\includegraphics[width=0.7\columnwidth]{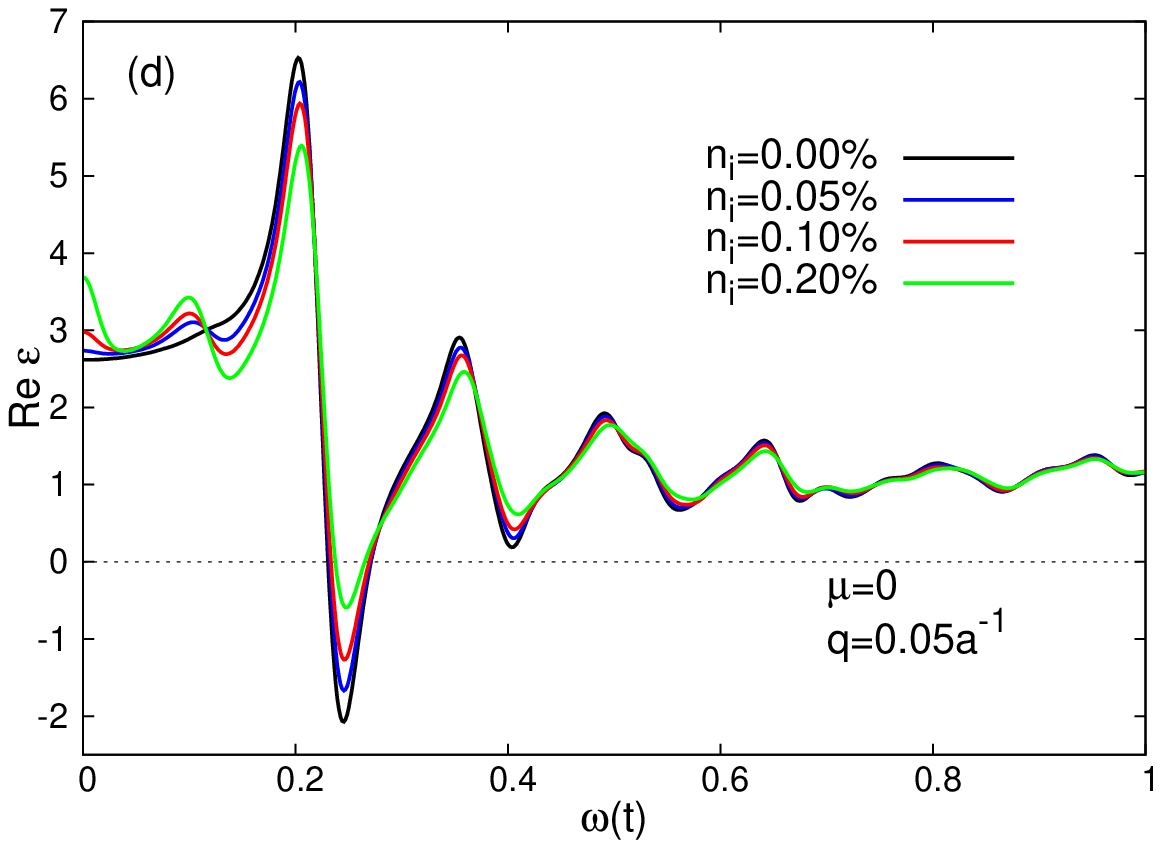}
}
\end{center}
\caption{Same as Fig. \protect\ref{Fig:PiClean} but in the presence of
resonant impurities. Panels (a) and (b) corresponds to $\mathrm{Im}~\Pi(q,%
\protect\omega)$ and $\mathrm{Re}~\protect\varepsilon(q,\protect\omega)$,
respectively, for a $\{10,6\}$ GAL with a random distribution of vacancies.
The different colors correspond to different amounts of missing dangling
bonds, as denoted in the inset of the figures. Panels (c) and (d)
corresponds to a GAL with hydrogen adatoms. Different colors correspond to
different percentage of adatoms in the sample. The insets in panels (a) and
(c) show the polarization in a broader range of energies.}
\label{Fig:PiResonant}
\end{figure*}

Therefore, the impurity bands which emerge in the spectrum due to the presence of
geometrical disorder, leads to a significant modification of the
electron-hole continuum as well as the dielectric function, and the results for GAL with this kind of disorder does not show any signature of plasmons. Nevertheless, notice that a resonance is still visible in the spectrum of a GAL
in which the size of the holes vary within some range, at an energy which is smaller than the energy of the plasmon mode in the clean GAL sample.
This is shown in Fig. \ref{Fig:DensityPlotDisorder}(b), where still a
pronounced resonance is observed at an energy $\Delta\approx 0.1t$.
The reason for low energy feature is clearly understood by
looking at the green dashed line of Fig. \ref{Fig:DOS}(d), which corresponds
to the DOS of a GAL sample with this kind of geometrical disorder, and where
we see that two new peaks have emerged in the DOS at energies $E\approx
\pm0.05t$. Therefore, the resonance in Fig. \ref{Fig:DensityPlotDisorder}(b)
at $\omega\approx 0.1t$ has its origin in electron-hole transitions between
those bands created in the spectrum by the effect of random radius
geometrical disorder. We emphasize that this is not a plasmon, since it does not correspond to a zero of the dielectric function.

We have also considered the effect of resonant impurities in the spectrum. This kind of disorder has a less dramatic effect on the excitation spectrum of the system, as it is shown in Fig. \ref{Fig:DensityPlotDisorder}. Both,
random distribution of vacancies and random distribution of hydrogen
impurities in the sample, have a similar effect on the dielectric function
of GAL, leading to a broadening of the plasmon modes, which are more
efficiently damped due to the presence of this kind of impurities. However, these features still correspond to a plasmon mode: this is proved by looking at the ${\rm Re}~\varepsilon$ plots of Fig. \ref{Fig:PiResonant} (b) and (d), which present well defined zeros for both, clean and disordered samples. Furthermore, the
energies of the gapped modes remain almost unchanged as compared to the spectrum of a perfect
and clean GAL [Fig. \ref{Fig:DensityPlot}(a)]. However, there is an additional transfer
of spectral weight to the low energy region of the spectrum, associated to
the impurity bands which have been created and which account for localized
states around the impurities.\cite{YRJK13} This feature in the spectrum is
especially visible in Fig. \ref{Fig:PiResonant}(a) and (c), where we observe a contribution to the polarization function in the low energy region of the spectrum which increases with the concentration of impurities in the sample.

\section{Static screening}

\label{Sec:Static}

\begin{figure*}[t]
\begin{center}
\mbox{
\includegraphics[width=0.85\columnwidth]{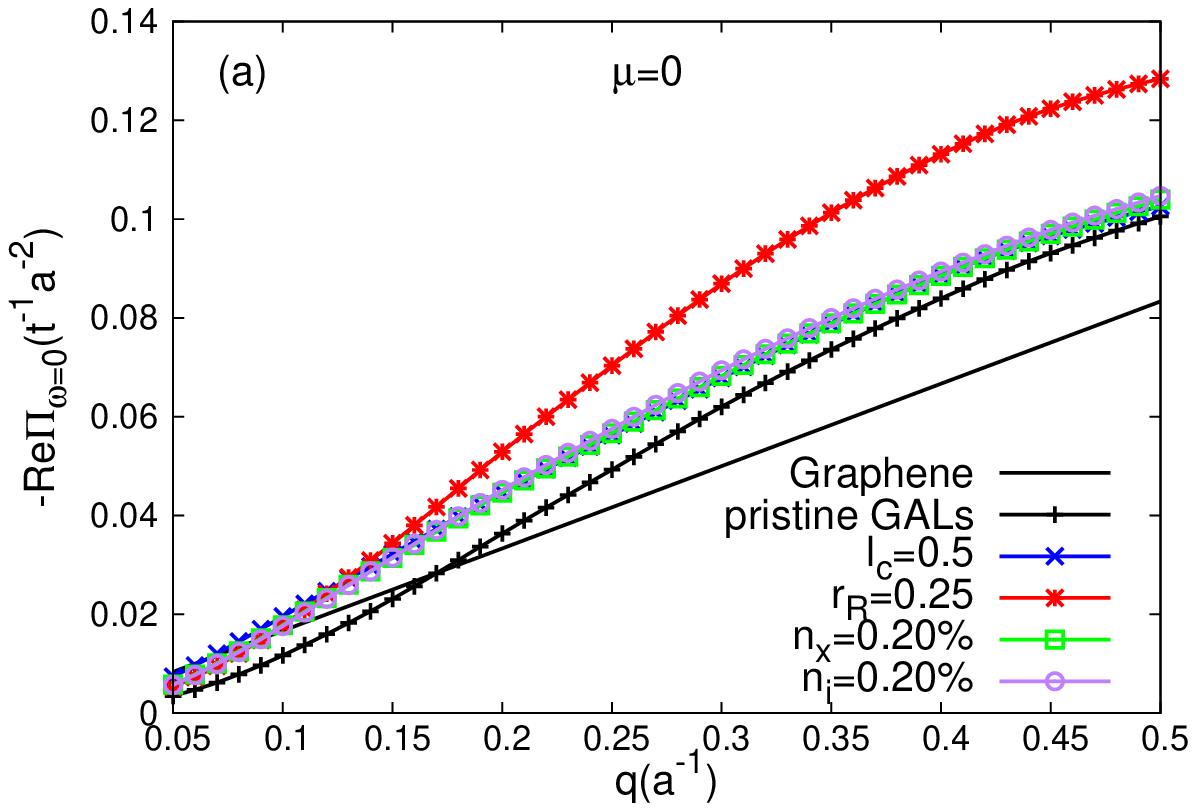}
\includegraphics[width=0.85\columnwidth]{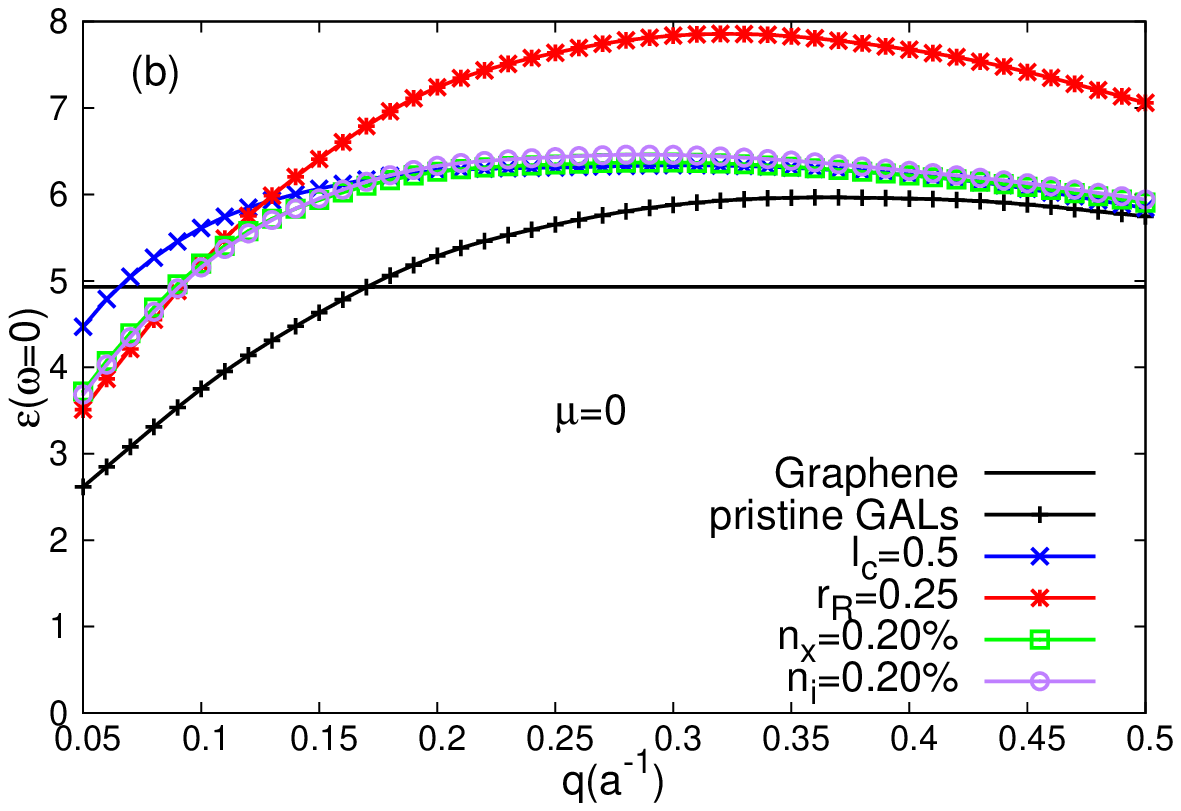}
}
\mbox{
\includegraphics[width=0.85\columnwidth]{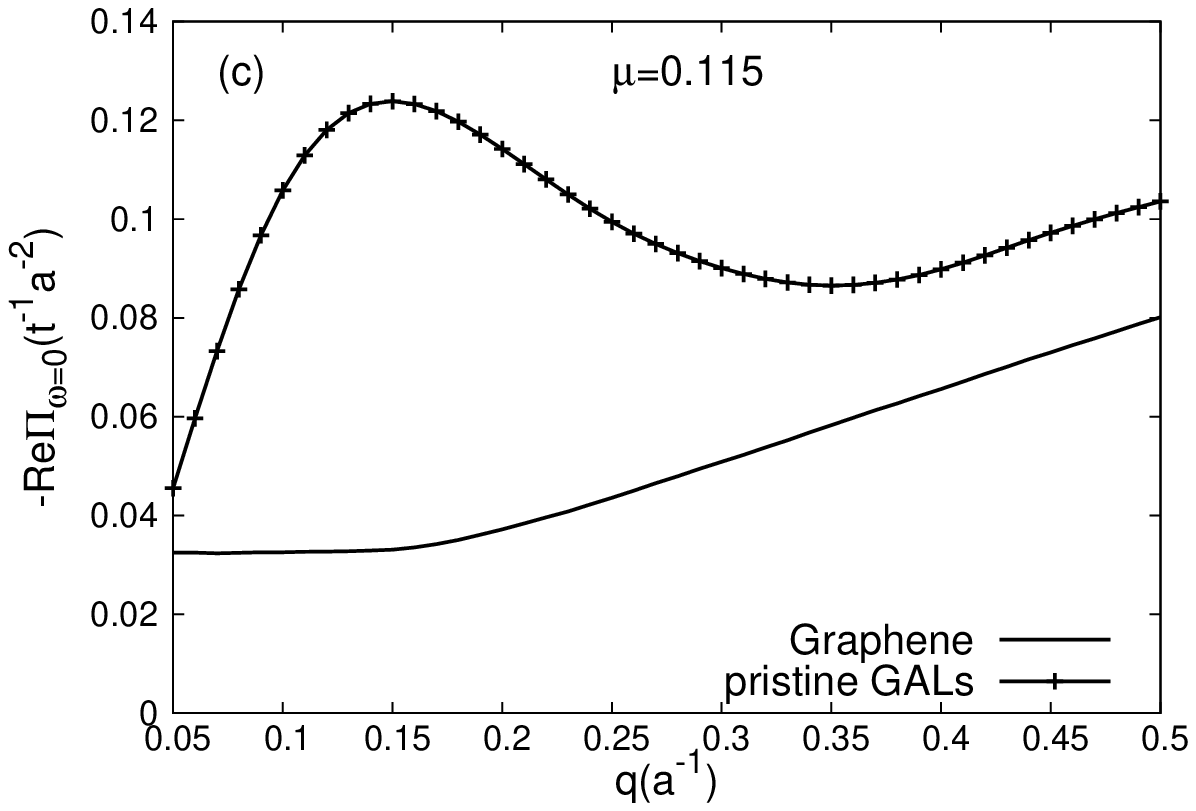}
\includegraphics[width=0.85\columnwidth]{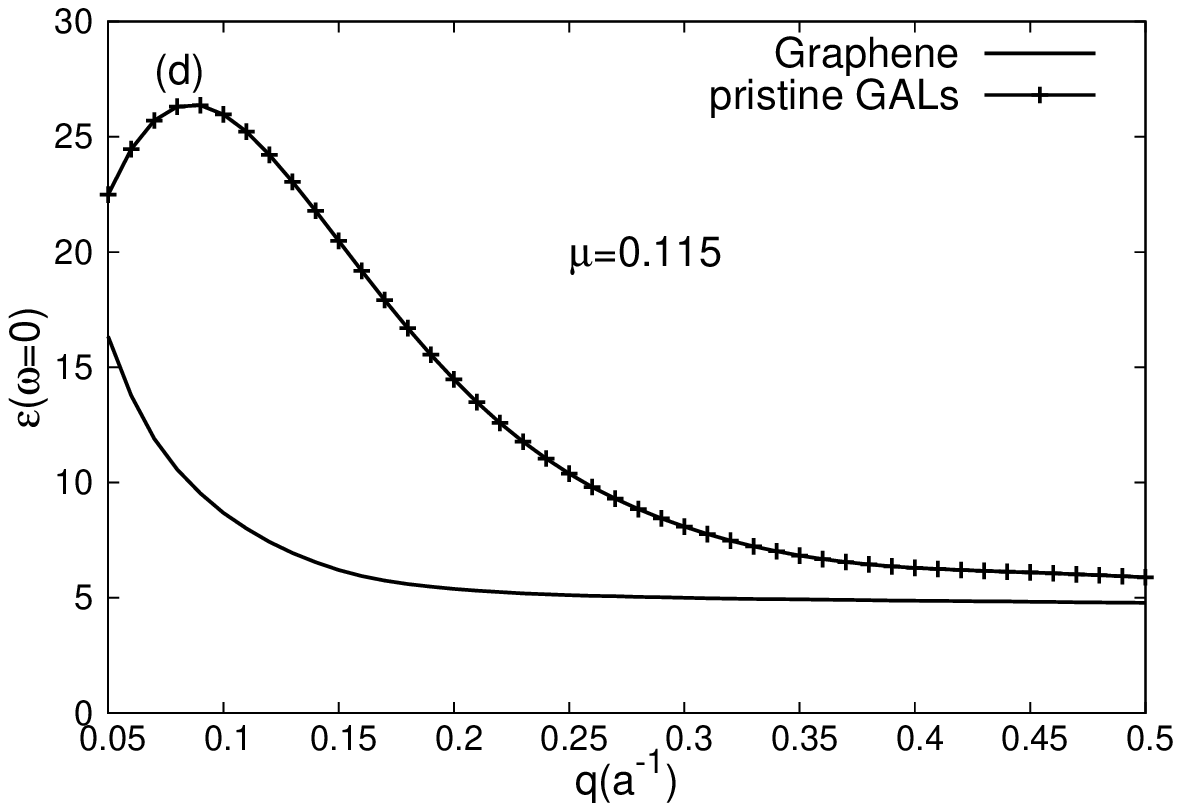}
}
\end{center}
\caption{Static polarization $\Pi(q,\protect\omega=0)$ and static dielectric
function $\protect\varepsilon(q,\protect\omega=0)$ of GALs. Plots (a) and
(b) are for undoped samples, $\protect\mu=0$, whereas plots (c) and (d)
correspond to a finite doping $\protect\mu=0.115t$. Different kinds of
disorder are considered in panels (a) and (b), as described in the inset.
For comparison, we include the corresponding polarization and dielectric
function for SLG, as given by the full black lines in each plot. }
\label{Fig:Static}
\end{figure*}

In this section we focus on the static dielectric screening of a GAL: we
calculate the polarization and dielectric function in the $\omega\rightarrow
0$ limit, using the Kubo formula Eq. (\ref{Eq:Kubo}). The results for clean
and disordered GALs, as compared to the corresponding polarization and
dielectric function for SLG, are shown in Fig. \ref{Fig:Static}. We start by
reviewing the main characteristics of the static screening in SLG. As is
well known,\cite{S86,GGV94,WSSG06,HS07,RB13} undoped SLG shows a linear increase
of the static polarization function with $q$, $\Pi_{\mathrm{inter}%
}(q)=-q/(4v_F)$, associated to inter-band transitions. Such contribution is
shown by the solid black line in Fig. \ref{Fig:Static}(a). For doped SLG,
there is an extra contribution due to intra-band excitations, which can be
written as
\begin{equation}
\Pi_{\mathrm{intra}}(q\le 2k_F)\approx -d(E_F)[1-q/(2k_F)]
\end{equation}
where $d(E_F)=E_F/(2\pi v_F^2)$ is the DOS at the Fermi level. Both
contributions lead to a constant static polarization function at small
wave-vectors, $\Pi(q\le 2k_F)=-d(E_F)$ as it is shown in Fig. \ref%
{Fig:Static}(c), typical of metallic screening,\cite{GV05} whereas the
inter-band linear term leads to an insulating like screening. The above
polarization function leads to a static dielectric function for SLG as shown
by the solid black lines of Fig. \ref{Fig:Static}(b), (d). Undoped SLG has a
constant dielectric function, $\varepsilon(q)=1+\pi e^2/(2\kappa
v_F)=\varepsilon_{\mathrm{SLG}}\approx 4.9$ for the parameters used in this
work, and it corresponds to the black horizontal line of Fig. \ref%
{Fig:Static}(b). Doping a SLG leads, as we have mentioned above, to a
metallic like screening with the corresponding $1/q$ divergence as $%
q\rightarrow 0$, as it is shown by the continuous black line in Fig. \ref%
{Fig:Static}(d).

The situation is different in GALs. First, Fig. \ref{Fig:Static}(a) shows a
static polarization function that grows with $q$, as in undoped SLG,
signaling a semiconducting type of screening. However, by comparing the
results of undoped GAL of Fig. \ref{Fig:Static}(a)-(b) to the corresponding
results of undoped SLG, we observe a transition from a region where $%
\varepsilon_{\mathrm{GAL}}(q)<\varepsilon_{\mathrm{SLG}}$, at long
wavelengths, to a region where $\varepsilon_{\mathrm{GAL}}(q)>\varepsilon_{%
\mathrm{SLG}}$ at shorter wavelengths. Therefore, we can expect a poorer
screening at long wavelengths in undoped GALs, as compared to undoped SLG,
whereas this tendency is inverted for shorter wavelengths, for which our
results suggest a more efficient screening in GALs than in SLG. The effect
of different kinds of disorder in the GAL tends to reinforce this tendency,
as we can see by corresponding curves for disordered GAL in Fig. \ref%
{Fig:Static}(a)-(b), although the qualitative behavior of the polarization
and dielectric function is not modified by disorder.

The differences in the static screening between GAL and SLG are even
stronger in the doped regime. As we can see in Fig. \ref{Fig:Static}(c)-(d),
$\Pi(q)$ for a GAL first grows linearly with $q$, until it reaches a maximum
for a characteristic wave-vector. This is clearly different to the
polarization function of doped SLG, which is constant with a magnitude given
by the DOS at the Fermi level. At large values of $q$, $\Pi(q)$ is again
linear with a slope similar to that of SLG, indicating a semiconductor-like
screening at large values of $q$. This difference is seen also in the
dielectric function of doped GALs, Fig. \ref{Fig:Static}(d), which does not
show the $q^{-1}$ divergence at $q\rightarrow 0$, representative of metallic
screening, but instead it presents a maximum at a given wavelength, and then
it decreases for smaller wave-vectors. Notice that no disorder is considered
in this calculation for doped GAL. We point out here that this behavior of
\textit{clean doped} GAL is similar to the polarization and dielectric
functions obtained for \textit{disordered undoped} SLG in the presence of
resonant scatterers.\cite{Yuan2012} In Ref. \onlinecite{Yuan2012} this bad
metal behavior in SLG below a characteristic length scale was identified
with Anderson localization. It is important to notice that, whereas the
localized states in SLG with resonant impurities are concentrated around the
impurities, in clean GAL the corresponding states are localized at the edges
of the antidots.\cite{Gunst2011,YRJK13} A more rigorous study should be made
in order to confirm if the reason for the dielectric properties observed
here for GALs might have the same origin as in chemically functionalized
graphene studied in Ref. \onlinecite{Yuan2012}.

\section{Conclusions}

\label{Sec:Conclusions}

In summary, we have performed a systematic study of the dielectric
properties of GALs. For this aim, we have used a tight-binding model in a
perforated honeycomb lattice of carbon atoms. The DOS, which has been
calculated from a numerical solution of the time-dependent Schr\"odinger
equation, shows a dramatic modification of the $\pi$ and $\pi^*$ bands of
SLG into a set of narrow and flat bands. We have further considered the most
generic sources of disorder in these kind of samples: geometrical disorder
such as random deviation of the periodicity and of the radii of the
nanoholes from the perfect array, as well as the effect of resonant
scatterers in the sample (e.g., vacancies, adatoms, etc.). The polarization
function has been obtained by using the Kubo formula for non-interacting
electrons, and electron-electron interactions have been considered within
the RPA. We have analyzed the main differences between the electron-hole
continuum, which limits the phase space available for particle-hole
excitations, of a GAL as compared to SLG. The conditions for the existence
of plasmon modes have been identified, and we find that damped and gapped
plasmons may exist in undoped GALs, associated to inter-band transitions
between the flat bands with a large DOS due the antidot array. Those modes
have a similar origin as the so-called $\pi$-plasmons in SLG, which are due
to inter-band transitions between states of the Van Hove singularities of $%
\pi$ and $\pi^*$ bands in SLG.\cite{EB08,YRK11} Furthermore, the
inter-subband plasmons in GALs are found to be almost dispersionless and, in
principle, they should be accessible by means of EELS experiments,\cite{EB08}
which could give information about the size of the gap opened in the sample.
For a doped GAL, when the chemical potential crosses one of the subbands in
the spectrum, we find that a classical plasmon with a dispersion $%
\omega(q)\propto \sqrt{q}$ is present. However, in a GAL the dispersion is
much weaker than for a SLG (in agreement with Ref. \onlinecite{Schultz2011}).

Finally, we studied the static screening in a GAL, by calculating the $%
\omega\rightarrow 0$ limit of the polarization and dielectric function.
Undoped GAL shows a mostly semiconductor-like screening, but with a much
rich structure as an undoped SLG. In the case of doped GAL, we find $%
\varepsilon(q)\propto q^{-1}$ only up to a characteristic wave-vector, for
which the dielectric function has a maximum after which it decays, showing a
bad metal behavior. The qualitative behavior found here for doped and clean
GAL is similar to that for an undoped disordered SLG in the presence of
resonant impurities.\cite{Yuan2012} In both cases, there is a characteristic
length scale which separates two different regimes in terms of screening.
However, whereas in the SLG systems studied in Ref. \onlinecite{Yuan2012}
the localized states are due to the presence of resonant scatterers, in the
present case are rather associated to localization at the edges of the
antidots.

\section{Acknowledgments}

The authors thank Jesper G. Pedersen for providing the band structure of GALs and Kristian S. Thygesen for useful remarks on the manuscript.
The support by the Netherlands National Computing Facilities foundation (NCF) and by the EC under the Graphene Flagship (Contract No. CNECT-ICT-604391) is
acknowledged. RR acknowledges financial support from the Juan de la Cierva Program and from
grant FIS2011- 23713 (MINECO, Spain). The Center for Nanostructured Graphene
(CNG) is sponsored by the Danish National Research Foundation, Project
DNRF58.

\bibliographystyle{apsrev}
\bibliography{BibliogrGrafene_APJ}

\end{document}